\documentclass[journal,twoside,romanappendices]{IEEEtran}

\usepackage{amsmath}
\usepackage{amssymb}
\usepackage{amsfonts}
\usepackage{amscd}
\usepackage{amsthm}
\usepackage[noadjust]{cite}
\usepackage{threeparttable}
\usepackage{textcomp}
\usepackage{multirow}

\usepackage{mathrsfs}
\usepackage{bbm}
\usepackage{amsbsy}

\usepackage{color}
\usepackage[dvipsnames]{xcolor}

\usepackage{paralist}

\usepackage[caption=false,font=footnotesize]{subfig}

\newtheorem{theorem}{Theorem}
\newtheorem{lemma}{Lemma}

\newtheorem{corollary}{Corollary}

\theoremstyle{definition}
\newtheorem{example}{Example}

\makeatletter
\renewcommand*\env@matrix[1][c]{\hskip -\arraycolsep
  \let\@ifnextchar\new@ifnextchar
  \array{*\c@MaxMatrixCols #1}}
\makeatother
\usepackage{graphicx}

\newcommand{\La}{\Lambda}
\newcommand{\Rb}{\mathbb{R}}
\newcommand{\Zb}{\mathbb{Z}}
\newcommand{\Lc}{\Lambda_{\rm c}}
\newcommand{\Bc}{\p{B}_{\rm c}}

\newcommand{\vol}{{\rm Vol}}
\newcommand{\vor}{\mathcal{V}}
\newcommand{\mmod}{{\rm~mod~}}
\newcommand{\p}{\pmb}
\newcommand{\rank}{{\rm rank}}
\newcommand{\rcov}{r_{\rm cov}}
\newcommand{\reff}{r_{\rm eff}}
\newcommand{\ball}{\mathcal{B}}
\newcommand{\Fp}{\mathbb{F}_p}
\newcommand{\Fb}{\mathbb{F}}
\newcommand{\Eb}{\mathbb{E}}
\newcommand{\snr}{{\sf SNR}}
\newcommand{\sig}{\sigma}
\newcommand{\sigz}{\sigma_{\p{z}}}
\newcommand{\al}{\alpha}
\newcommand{\lc}{\p{\lambda}_{\rm c}}
\newcommand{\tr}{\intercal}
\newcommand{\Pp}{{\rm P}}
\newcommand{\CS}{\mathscr{C}_{\p{S}}}
\newcommand{\LS}{\Lambda_{\p{S}}}

\newcommand{\rz}{r_{\p{z}}}
\newcommand{\ballz}{\mathcal{B}_{r_{\p{z}}}}
\newcommand{\maxflow}{\operatorname{max-flow}}

\title{Lattice Codes Achieve the Capacity of Common Message Gaussian Broadcast Channels with Coded Side Information}
\author{
\IEEEauthorblockN{Lakshmi~Natarajan, Yi~Hong,~\IEEEmembership{Senior~Member,~IEEE}, and Emanuele~Viterbo,~\IEEEmembership{Fellow,~IEEE}}%
\thanks{%Manuscript received September~03,~2015; revised January~02,~2017, August~21,~2017, November~21,~2017; accepted November~22,~2017.% 
This work was supported by the Australian Research Council (ARC) Discovery Project (ARC~DP160101077). The material in this paper was presented in part at the IEEE International Symposium on Information Theory, Aachen, Germany, June~25--30, 2017.}%
\thanks{Dr. Natarajan is with the Department of Electrical Engineering, Indian Institute of Technology Hyderabad, Sangareddy 502\,285, India (e-mail: lakshminatarajan@iith.ac.in).}%
\thanks{Dr. Hong and Prof. Viterbo are with the Department of Electrical and Computer System Engineering, Monash University, VIC 3800, Australia (e-mail: \{yi.hong, emanuele.viterbo\}@monash.edu).}%
}

\begin{document}

%% \IEEEpubid{0000--0000/00\$00.00~\copyright~2017 IEEE}

\maketitle

\begin{abstract}
\boldmath
Lattices possess elegant mathematical properties which have been previously used in the literature to show that structured codes can be efficient in a variety of communication scenarios, including coding for the additive white Gaussian noise (AWGN) channel, dirty-paper channel, Wyner-Ziv coding, coding for relay networks and so forth.
We consider the family of single-transmitter multiple-receiver Gaussian channels where the source transmits a set of common messages to all the receivers { (multicast scenario)}, and each receiver has \emph{coded side information}, i.e., prior information in the form of linear combinations of the messages.
This channel model is motivated by applications to multi-terminal networks where the nodes may have access to coded versions of the messages from previous signal hops or through orthogonal channels. 
The capacity of this channel is known and follows from the work of Tuncel~(2006), which is based on random coding arguments.
In this paper, following the approach of Erez and Zamir, we design lattice codes for this family of channels when the source messages are symbols from a finite field $\Fb_p$ of prime size. Our coding scheme utilizes Construction~A lattices designed over the same prime field $\Fb_p$, and uses \emph{algebraic binning} at the decoders to expurgate the channel code and obtain good lattice subcodes, for every possible set of linear combinations available as side information. The achievable rate of our coding scheme is a function of the size $p$ of underlying prime field, and approaches the capacity as $p$ tends to infinity.
\end{abstract}


\section{Introduction} \label{sec:introd}

\IEEEPARstart{I}{nformation}-theoretic results often rely on random coding arguments to prove the existence of good codes. 
Usually, the codebook is constructed by randomly choosing the components of each codeword independently and identically from a judiciously chosen probability distribution. While this technique is powerful, the resulting codebooks do not exhibit any structure that may be of practical interest. One such desirable structure is linearity, which allows complexity reductions at the encoder and decoder by utilizing efficient algebraic processing techniques. 
{%%
Further, in certain communication scenarios, coding schemes based on linear codes yield a larger achievable rate region than random code ensembles, as was shown by K{\"o}rner and Marton~\cite{KoM_IT_79} for a distributed source coding problem. 
}
Structured coding schemes have been widely studied in the literature, especially for communications in the presence of side information and in multi-terminal networks. 
For an overview of structured coding schemes we refer the reader to~\cite{ZSE_IEEE_IT_02,NaG_ETT_08} and the references therein.

For communication in the wireless domain, structured codes can be obtained by choosing finite subsets of points from lattices~\cite{For_IEEE_IT_88,ZSE_IEEE_IT_02,CoS_Springer_99,deB_IT_75}. 
A lattice is an infinite discrete set of points in the Euclidean space that are regularly arranged and are closed under addition. 
Codes based on lattices, known as \emph{(nested) lattice codes} or \emph{Voronoi codes}, are the analogues of linear codes in wireless communications.
Efficient lattice based strategies are known for a variety of communication scenarios, such as for achieving the capacity of the point-to-point additive white Gaussian noise (AWGN) channel~\cite{UrR_IT_98,ErZ_IEEE_IT_04,OrE_IT_16,diP_thesis_14,LiB_IT_14}, for dirty-paper coding~\cite{ESZ_IT_05,ZSE_IEEE_IT_02}, the Wyner--Ziv problem~\cite{ZSE_IEEE_IT_02} and communication in relay networks~\cite{NCL_IT_10,WNPS_IT_10,NaG_IT_11,FSK_IT_13}, to name only a few.

%% \IEEEpubidadjcol

{%%
In this paper we present {good} lattice strategies for communication in common message Gaussian broadcast channels, which we refer to as the {\em multicast channel}, where receivers have prior side information about the messages being transmitted. 
}%
In particular, we assume that the transmitter is { multicasting} $K$ message symbols $w_1,\dots,w_K$ from a finite field $\Fp$, of prime size $p$, to all the receivers, and each receiver may have \emph{coded side information} about the messages: the prior knowledge of the values of (possibly multiple) $\Fp$-linear combinations of $w_1,\dots,w_K$. The number of linear combinations available as side information and the coefficients of these linear combinations can differ from one receiver to the next.  
{%%
The capacity of this channel is known and follows from the results of Tuncel~\cite{Tun_IEEE_IT_06}, where the achievability part utilizes an ensemble of codebooks generated using the Gaussian distribution.
}

{
The multiuser channel considered in this paper is a noisy version of a simple special case of \emph{index coding}~\cite{YBJK_IEEE_IT_11,ALSWH_FOCS_08,RSG_IEEE_IT_10}. The index coding problem considers a {\em noiseless} broadcast link where each receiver demands a subset of the source messages and knows the values of some other subset as side information. 
A generalization of the index coding problem in which the receivers have access to linear combinations of messages was studied recently in~\cite{SDS_PIMRC_12,LDH_COMML_15}.
The specific instance of index coding where each receiver demands all the messages from the source corresponds to a noiseless multicast channel and has a simple optimum solution based on maximum distance separable (MDS) codes~\cite{BiK_INFOCOM_98}.
When the channel is noisy, capacity-achieving coding schemes based on structured codes are not available. In this paper we design lattice-based strategies for multicasting over the AWGN channel where the side information at the receivers is in the form of linear combinations of source messages.
}%

The case of Gaussian { multicast} channel with coded side information is motivated by applications to multi-terminal communication networks. It is known that signal interference in wireless channels can be harnessed by decoding linear combinations of transmit messages instead of either treating interference as noise or decoding interference along with the intended message~\cite{NaG_IT_11,WNPS_IT_10}. When such a technique is used in a mutli-hop communication protocol, one encounters receivers that have coded side information obtained from transmissions in the previous phases. Similarly, in a network that consists of both wired and wireless channels, the symbols received from wired links can be utilized as side information for decoding the wireless signals. If a linear network code is used in the wired part of the network, then the side information is in the form of linear combinations of the source messages. 

\begin{example}[\emph{Communciation in relay networks}] \label{ex:relay_network}
\begin{figure}[t!]
\centering
\subfloat[Multiple-access phase: The relay ${\sf R}$ decodes both $w_1$ and $w_2$, while ${\sf U}_1,{\sf U}_2,{\sf U}_3$ decode $w_1$,$w_2$ and $s_1w_1 + s_2w_2$, respectively.]{\includegraphics[width=3in]{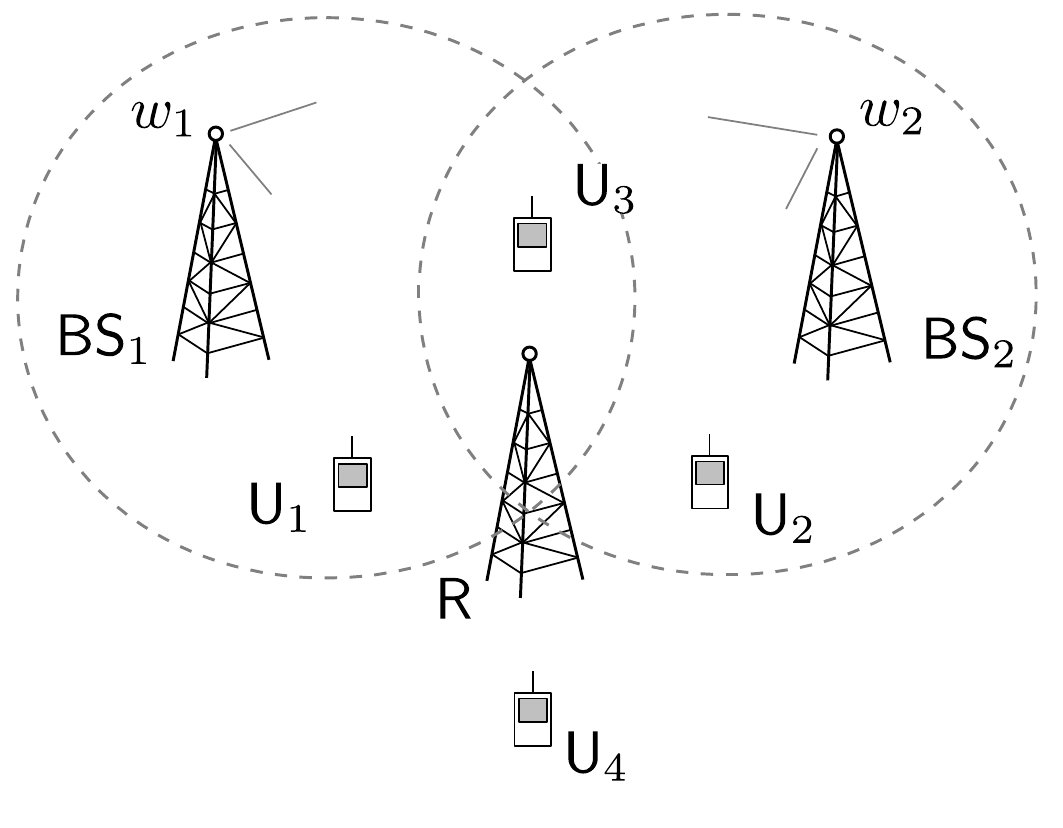}}
\hfil
\subfloat[{ Multicast} phase: ${\sf R}$ { multicasts} $w_1$ and $w_2$ to all four user nodes. Three of the users have the knowledge of some linear combination of $w_1$ and $w_2$, while the fourth user has no side information.]{\includegraphics[width=3in]{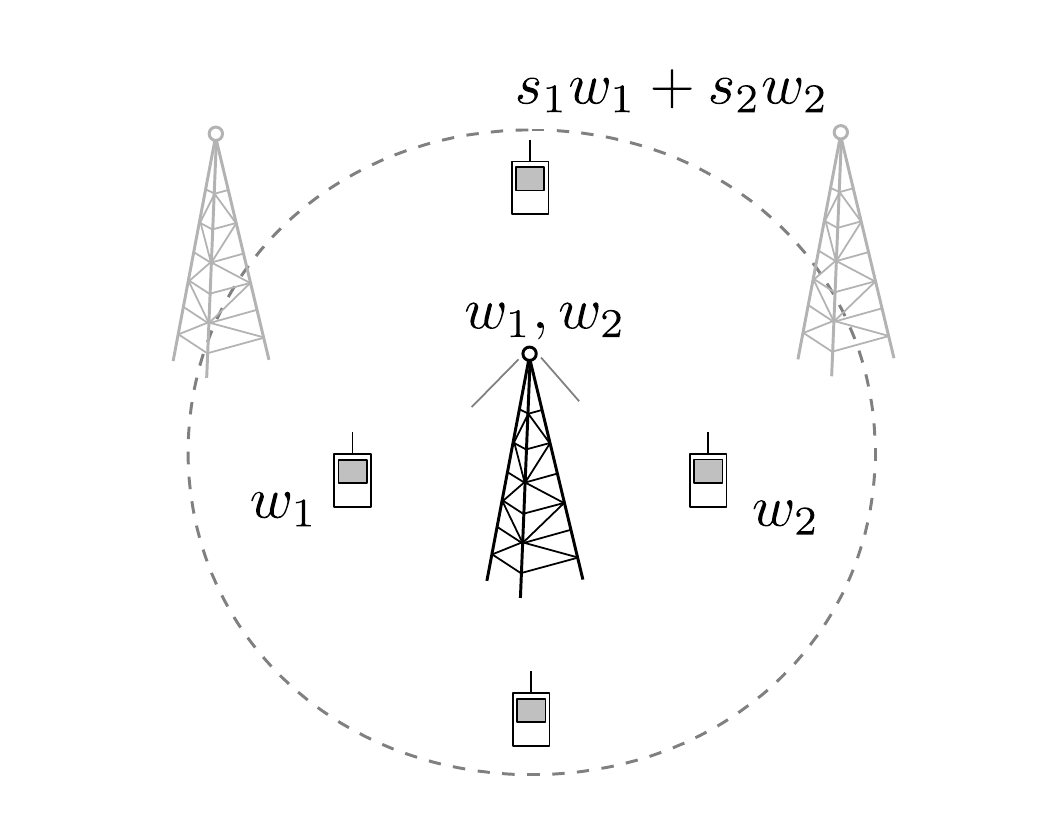}}
\vspace{2mm}
\caption{A relay network where one encounters a { common message broadcast channel} with coded side information at the receivers.}
\label{fig:relay_network}
\end{figure}
Consider a wireless network with two base stations ${\sf BS}_1$ and ${\sf BS}_2$, that hold message symbols $w_1$ and $w_2$, respectively. The base stations are required to { multicast} $w_1$ and $w_2$ to four user nodes ${\sf U}_1,\dots,{\sf U}_4$ through the relay node ${\sf R}$, see Fig.~\ref{fig:relay_network}. 
In the first phase of the protocol, ${\sf BS}_1$ and ${\sf BS}_2$ encode the data symbols $w_1$ and $w_2$, and transmit the resulting codewords simultaneously. 
By using the decoding technique of compute-and-forward~\cite{NaG_IT_11}, ${\sf U}_3$ reliably decodes some linear combination $s_1w_1+s_2w_2$, $s_1,s_2 \in \Fp$, from the received noisy superposition of the two transmit signals.
On the other hand, ${\sf R}$ has a higher signal-to-noise ratio and successfully decodes both $w_1$ and $w_2$ by behaving as a multiple-access receiver.
Further, there is no signal interference at ${\sf U}_1$ and ${\sf U}_2$, and these two nodes reliably decode $w_1$ and $w_2$, respectively.

We observe that the second phase of the protocol is a { common message broadcast channel} with coded side information at the receivers: the relay needs to { multicast} $w_1,w_2$ to four user nodes, the first three users ${\sf U}_1,{\sf U}_2,{\sf U}_3$ have prior knowledge of the linear combinations $w_1+0w_2$, $0w_1+w_2$ and $s_1w_1+s_2w_2$, respectively, while the fourth user has no such side information.
\hfill \IEEEQED
\end{example}

\begin{example}[\emph{Wireless overlay for wired networks}] \label{ex:overlay}
Assume a network of noiseless wired links in the form of a directed acyclic graph, where the source node $v_{\sf s}$ desires to multicast $K$ independent messages $w_1,\dots,w_K \in \Fp$ to a set of destination nodes $\mathcal{D}$. 
The wireline network employs a traditional \emph{(scalar) linear network code}~\cite{LYC_IT_03,KoM_IT_03,Yeu_Springer_08}, i.e., the symbol transmitted on each outgoing edge of a node is an element of $\Fp$ generated as a linear combination of the symbols received on its incoming edges. At every destination node $v_{\sf d} \in \mathcal{D}$, the decoder attempts to recover the $K$ message symbols from their $\Fp$-linear combinations received on its incoming edges. Recovery is possible if and only if the number of linearly independent equations available at $v_{\sf d}$ is $K$. 
It is known that the maximum number of linearly-independent equations that can be made available at $v_{\sf d}$ is $\min\{\maxflow(v_{\sf d}),K\}$, where $\maxflow(v_{\sf d})$ is the maximum number of edge-disjoint paths from $v_{\sf s}$ to $v_{\sf d}$, see~\cite{Yeu_Springer_08}. 
It follows that multicasting is possible if and only if $\maxflow(v_{\sf d}) \geq K$ for every $v_{\sf d} \in \mathcal{D}$.

Now suppose there exist destination nodes with $\maxflow$ less than $K$, i.e., the communication demands are beyond the wireline network's capacity. 
A solution to meet the demands is to broadcast a wireless signal from the source to fill the capacity deficiency of the wired network, see Fig.~\ref{fig:overlay}.
At each destination, the $\Fp$-linear combinations obtained from the wireline network serve as side information to decode the wireless broadcast signal. 
\hfill\IEEEQED
\begin{figure}[!t]
\centering
\includegraphics[width=3in]{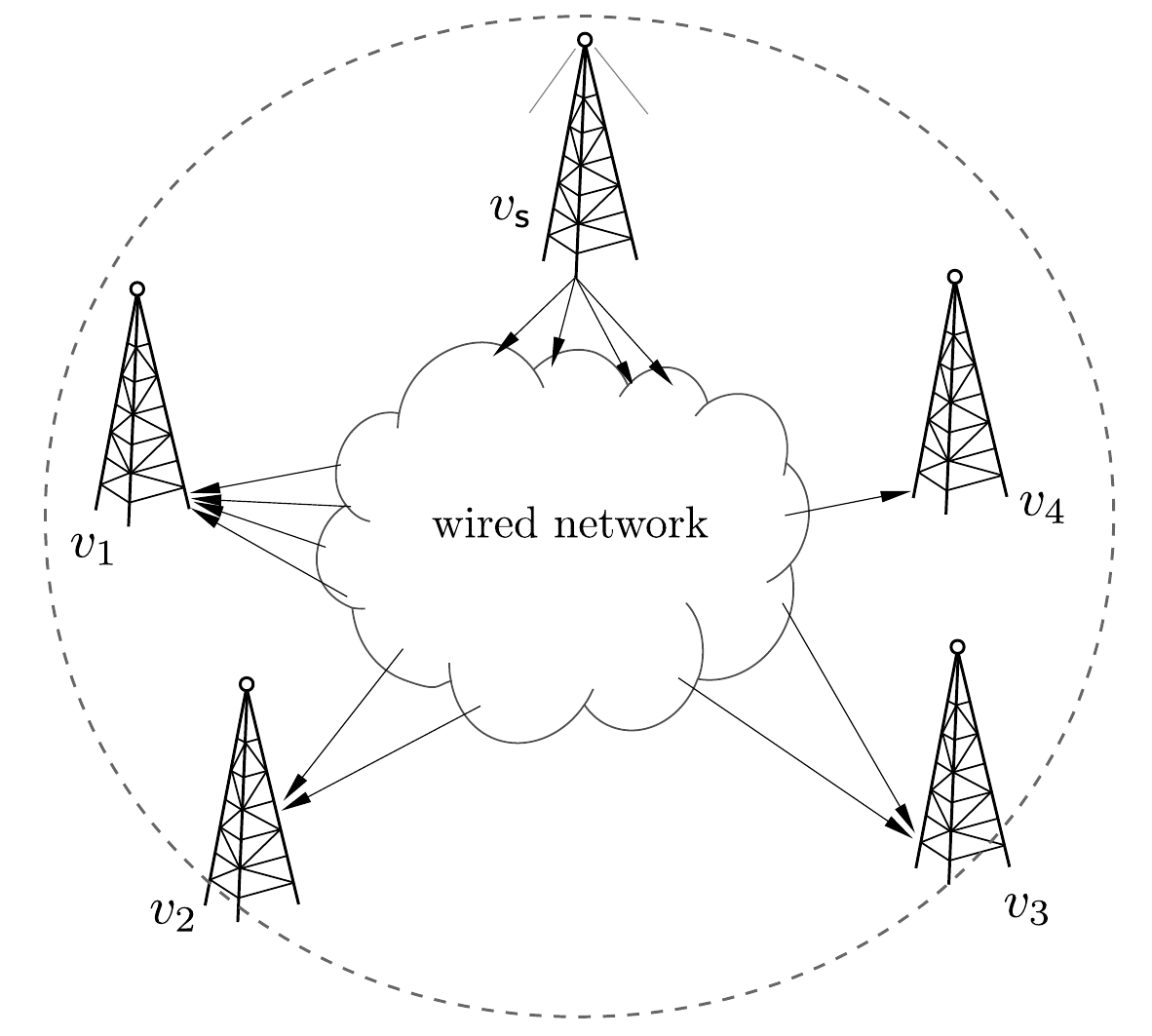}
\caption{Wireless { multicast} as an overlay for a wired network: the source node $v_{\sf s}$ encodes $w_1,\dots,w_K$ and broadcasts wirelessly to the destination nodes $v_1,\dots,v_4$ in order to supplement the communication through the wired network. Each destination receives (possibly multiple) $\Fp$-linear combinations of the source symbols through a linear (wireline) network code and uses this information to decode the wireless signal broadcast by $v_{\sf s}$.}
\label{fig:overlay}
\end{figure} 
\end{example}

A special case of coded side information is the Gaussian { multicast} channel where each receiver has prior knowledge of the values of some subset of the $K$ messages. The known capacity-achieving coding schemes for this special case are based on random coding using i.i.d. (independent and identically distributed) codewords~\cite{Tun_IEEE_IT_06,Xie_CWIT_07,KrS_ITW_07,Wu_ISIT_07,OSBB_IEEE_IT_08,XuS_ITW_07}.
Existence of lattice based capacity-achieving coding schemes were proved in~\cite{KrS_ITW_07,WLS_arxiv_15} for the special case where the number of messages and receivers are two and each receiver has the knowledge of one of the messages.  
Constructions of binary codes for this channel were proposed in~\cite{XFKC_CISS_06,BaC_ITW_11,MLV_PIMRC_12}.
Explicit constructions of lattice codes were given in~\cite{NHV_IT_15,Hua_ISIT_15} that convert receiver side information into additional apparent coding gain in the AWGN channel.
Codes based on quadrature amplitude modulation were constructed in~\cite{NHV_Comml_15,NHV_ISIT_15}.
In~\cite{WLS_arxiv_15}, explicit codes based on lattices and coded modulation have been designed that perform within a few decibels of capacity when the number of receivers is two and each knows one of the two messages being transmitted.

The objective of this paper is to prove that lattice codes can achieve the capacity of the {common message Gaussian broadcast channels} with coded side information. 
We use the information-theoretic framework set by Erez and Zamir~\cite{ErZ_IEEE_IT_04} to this end.
{The proposed coding scheme uses lattices obtained by applying Construction~A to linear codes over the prime field $\Fb_p$ which is the alphabet of the source messages. The achievable rate of our lattice-based coding scheme is a function of the prime $p$, and approaches the capacity of the common message Gaussian broadcast channel as $p \to \infty$.}

Our decoding scheme involves \emph{algebraic binning}~\cite{ZSE_IEEE_IT_02} where the receiver side information is used to expurgate the channel code and obtain a lower rate subcode.
The set of linear equations available as side information may differ from one receiver to another, and hence, each receiver must employ a different binning scheme for the same channel code.
The coding scheme ensures that the binning performed at each receiver produces a good lattice subcode of the transmitted code. 
Following expurgation, each receiver decodes the channel output by minimum mean square error (MMSE) scaling and quantization to an infinite lattice.
The algebraic structure of the coding scheme facilitates the performance analysis by decomposing the original channel into multiple independent point-to-point AWGN channels -- one corresponding to each receiver -- where each of the point-to-point AWGN channels uses a lattice code for communication.
Unlike~\cite{ErZ_IEEE_IT_04}, where achievability in a point-to-point AWGN channel was proved using error exponent analysis, we provide a direct proof based only on simple counting arguments. 
  
As a corollary to the main result, we obtain an alternative proof of the goodness of lattice codes in achieving the capacity of the point-to-point AWGN channel. 
Previous proofs of this result presented in~\cite{ErZ_IEEE_IT_04,OrE_IT_16,diP_thesis_14} also use ensembles of lattices obtained by applying Construction~A to random linear codes over a prime field $\Fp$; see also~\cite{Loe_IT_97,KrP_Good_Lattices_07}. While~\cite{ErZ_IEEE_IT_04} used primes $p$ that were exponential in the code length $n$,~\cite{OrE_IT_16} and~\cite{diP_thesis_14} improved this result to let $p$ grow as $n^{1.5}$ and $n^{0.5}$, respectively. 
{%%
The corollary presented in this paper further improves these results by enabling a choice of the prime $p$ which is independent of the code length $n$ but is a function only of the gap between the desired rate and the channel capacity.
}

{ 
Lattices have been used to design powerful physical-layer coding schemes for wireless networks consisting of multiple sources, relays and destinations~\cite{NaG_IT_11,WNPS_IT_10,NCL_IT_10,FSK_IT_13}.
In these networks information from the source nodes is conveyed to the destination nodes through relays over multiple hops and time slots. In each time slot, a set of nodes act as transmitters and every other node in their range observes a linear superposition of the transmitted signals perturbed by AWGN. 
Lattice coding schemes for these networks are designed such that each receiver can reliably decode the observed noisy superposition to a linear combination of source messages which it then proceeds to transmit in the next time slot. Every destination node decodes its desired messages once it collects sufficiently many linear combinations.
In contrast, in this paper we consider a single hop interference-free transmission in a multicast channel consisting of one transmitter and multiple destination nodes that are aided by coded side information. Our objective is to design coding schemes that can utilize prior knowledge at these receivers rather than exploit wireless interference arising from multiple simultaneous transmissions, as often experienced in relay networks.
}%

The organization of this paper is as follows.
We introduce the channel model in Section~\ref{sec:channel_model} and review the relevant background on lattices and lattice codes in Section~\ref{sec:lattice_prelim}. In Section~\ref{sec:3}, we state the main theorem, and describe the lattice code ensemble and encoding and decoding procedures. We prove the main theorem and state a few corollaries in Section~\ref{sec:proof_of_main}, and finally, we discuss some concluding remarks in Section~\ref{sec:conclusion}.

{\it Notation:} Matrices and column vectors are denoted by bold upper and lower case letters, respectively. The symbol $\|\cdot\|$ denotes the Euclidean norm of a vector, and $(\cdot)^\tr$ is the transpose of a matrix or a vector. 
The Kronecker product of two matrices $\p{A}$ and $\p{B}$ is $\p{A} \otimes \p{B}$, $\p{I}_{\ell}$ is the $\ell \times \ell$ identity matrix, and $\p{0}$ is the all zero matrix of appropriate dimension.
The symbol $\log(\cdot)$ denotes logarithm to the base $2$ and ${\rm ln}(\cdot)$ denotes logarithm to the base $e$.
The expectation operator is denoted by $\Eb$. The symbol $\mathcal{M} \backslash \mathcal{N}$ denotes the elements in the set $\mathcal{M}$ that do not belong to the set $\mathcal{N}$.

\section{Channel Model and Lattice Preliminaries} \label{sec:2}

\subsection{Channel Model and Problem Statement} \label{sec:channel_model}

We consider a (non-fading) { common message Gaussian broadcast channel} with a single transmitter and finitely many receivers, where all terminals are equipped with single antennas. The transmitter operates under an average power constraint and the receivers are affected by additive white Gaussian noise with possibly different noise powers.
There are $K$ independent messages $w_1,\dots,w_K$ at the transmitter that assume values with a uniform probability distribution from a prime finite field $\Fp$. Each receiver desires to decode all the $K$ messages while having prior knowledge of the values of some $\Fp$-linear combinations of the messages $w_1,\dots,w_K$. 
Consider a generic receiver that has access to the values $u_m$, $m=1,\dots,M$, of the following set of $M$ linear equations
\begin{equation*}
\sum_{k=1}^{K} s_{m,k}w_k = u_m,~~~~ m=1,\dots,M.
\end{equation*} 
We will denote this side information configuration using the matrix $\p{S} = [s_{m,k}] \in \Fp^{M \times K}$, where each row of $\p{S}$ represents one linear equation. Any row of $\p{S}$ that is linearly dependent on the other rows represents redundant information and can be discarded with no loss to the receiver side information, and hence, with no loss to system performance. Hence, without loss in generality, we will assume that the rows of $\p{S}$ are linearly independent over $\Fp$, i.e., $\rank(\p{S})=M$, and $M<K$. Note that the values of $\p{S}$ and $M$ can be different across the receivers. A receiver with no side information is represented with an empty matrix for $\p{S}$ (with $M=0$).

A receiver in the { multicast channel} is completely characterized by its \emph{(coded) side information matrix} $\p{S}$ and the variance $\sig^2$ of the additive noise. If we assume that the average transmit power at the source is $1$, then the signal-to-noise ratio at this receiver is $\snr=\frac{1}{\sig^2}$. We will denote a receiver by the pair $(\p{S},\sig^2)$, where $\p{S}$ is any matrix over $\Fp$ with $K$ columns and linearly independent rows, and $\sig^2>0$. 
Note that \emph{uncoded} side information, i.e., the prior knowledge of the values of a size $M$ subset of $w_1,\dots,w_K$, is a special case, and hence, is contained within the definition of our channel model.

\begin{example}
Consider a source transmitting $K=3$ symbols, $w_1,w_2,w_3$, from the finite field $\Fb_5=\{0,1,2,3,4\}$. A receiver that has prior knowledge of the value of $w_2$ has side information matrix $\p{S} = \begin{pmatrix} 0 & 1 & 0 \end{pmatrix}$. This corresponds to the equation $0w_1 + 1w_2 + 0w_3$, and the number of linearly independent equations at this receiver is $M=\rank(\p{S})=1$. 

Now consider another receiver that has the knowledge of the values of the following three equations: $w_1+4w_2+3w_3$, $4w_1 + 3w_2$ and $2w_1 + w_2 + 3w_3$. In matrix form, this side information is represented by
\begin{equation*}
 \begin{pmatrix} 1 & 4 & 3 \\ 4 & 3 & 0 \\ 2 & 1 & 3 \end{pmatrix},
\end{equation*} 
where the three rows represent the three equations, in that order. The first row of this matrix is equal to the sum (over $\Fb_5$) of the second and third rows, and hence, the side information from the first equation is redundant and can be discarded. Since the remaining two rows are linearly independent, the side information at this receiver can be represented by the following matrix that consists of these two rows,
\begin{equation*} %\label{eq:ex:side_inf:1}
 \p{S}= \begin{pmatrix} 4 & 3 & 0 \\ 2 & 1 & 3 \end{pmatrix}.
\end{equation*} 
The number of linearly independent equations at this receiver is $M=\rank(\p{S})=2$. 
\hfill\IEEEQED
\end{example}

\begin{figure*}[!t]
\centering
\includegraphics[width=5in]{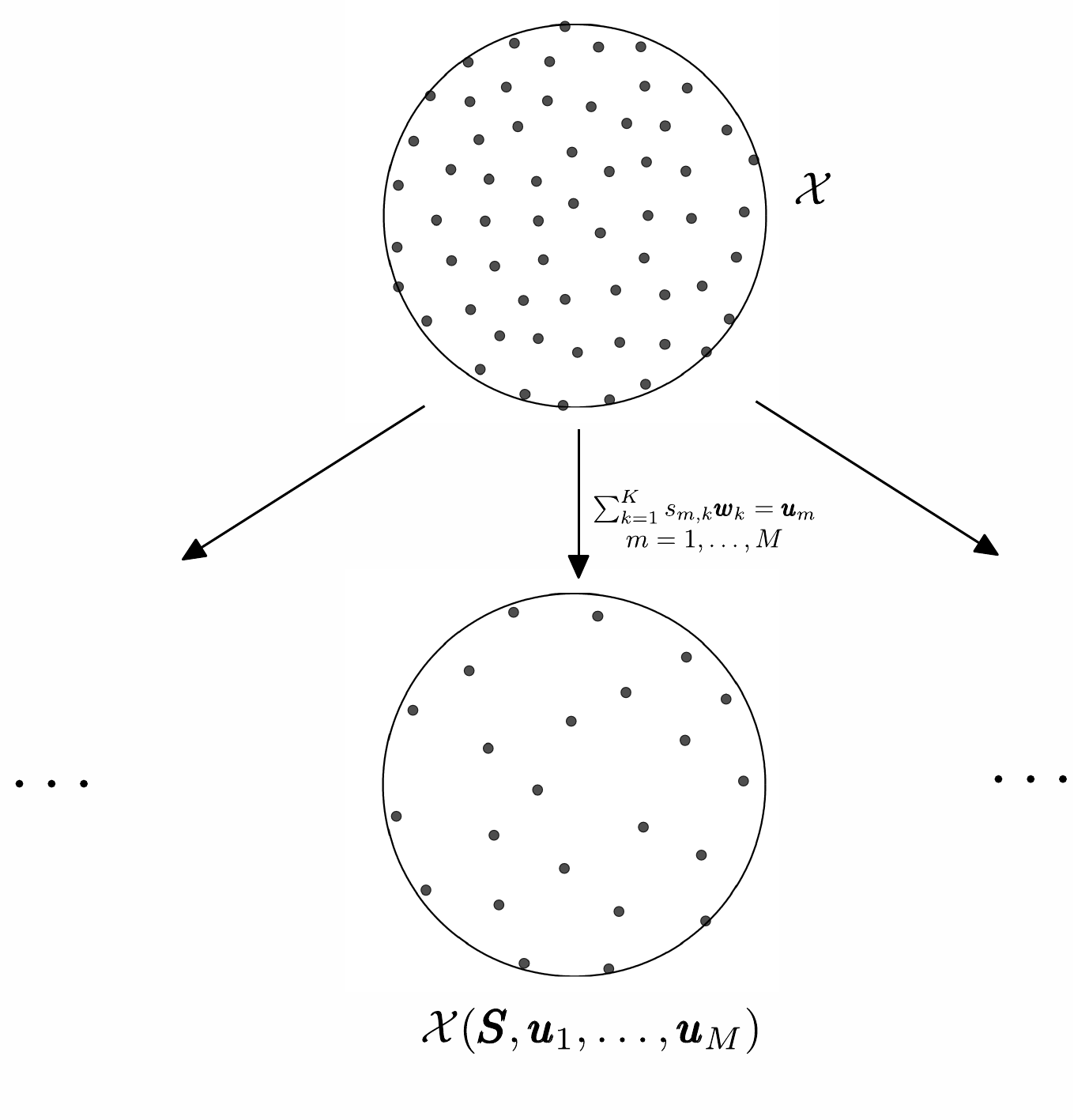}
\caption{Each receiver $(\p{S},\sig^2)$ of the { multicast channel} uses its own side information $(\p{S},\p{u}_1,\dots,\p{u}_M)$ to expurgate the channel code $\mathcal{X}$ and obtain a subcode $\mathcal{X}(\p{S},\p{u}_1,\dots,\p{u}_M)$. Note that the resulting subcodes can be different across the receivers. In order to achieve the capacity of the { multicast channel}, we require that each of these expurgated codes be good for channel coding over the point-to-point AWGN channel.
}
\label{fig:expurg}
\end{figure*} 

From elementary linear algebra we know that if the values $u_m$ of $M$ linearly independent combinations of the variables $w_1,\dots,w_K$ are given, then the set of all possible solutions of $(w_1,\dots,w_K)$ is a coset of a $(K-M)$ dimensional linear subspace of $\Fp^K$. Since the a~priori probability distribution of $w_1,\dots,w_K$ is uniform, we conclude that, given the side information values $u_m$, $m=1,\dots,M$, the probability distribution of $(w_1,\dots,w_K)$ is uniform over this coset. Using the fact that the number of elements in the coset is $p^{K-M}$, we observe that the conditional entropy of $(w_1,\dots,w_K)$ given the side information is
\begin{align} \label{eq:conditional_entropy}
 H\left(w_1,\dots,w_K \, \vert \, u_1,\dots,u_M \right) &= \log \left( p^{K-M} \right) \nonumber \\ &= (K-M) \log p.
\end{align} 
Suppose we want to transmit, on the average, one realization of $(w_1,\dots,w_K)$ in every $\kappa$ uses of the broadcast channel. The transmission rate of each message is $R=\frac{1}{\kappa} \log p$~b/dim (bits per real dimension or bits per real channel use). 

For the simplicity of exposition, we consider only the symmetric case where all the $K$ messages are required to be transmitted at the same rate $R$. The general scenario, where the messages are of different rates, can be reduced to the symmetric case through \emph{rate-splitting}: if there are $K'$ messages with transmission rates $r_1,\dots,r_{K'}$, respectively, then by splitting each of these original sources into multiple virtual sources, one can generate a set of $K$ sources ($K \geq K'$) such that their rates $R_1,\dots,R_K$ are as close to each other as required.

We will assume that the encoding at the transmitter is performed on a block of $\ell$ independent realizations of the $K$ message symbols, i.e., the source jointly encodes $K$ message vectors $\p{w}_1,\dots,\p{w}_K \in \Fp^\ell$.
The transmitter uses an $n$-dimensional channel code $\mathcal{X} \subset \Rb^n$ together with a function 
\begin{equation*}
\rho:\Fp^\ell \times \cdots \times \Fp^\ell \to \mathcal{X}
\end{equation*} 
to jointly encode the $K$ message vectors. 
The number of codewords in $\mathcal{X}$ is $p^{K\ell}$, and we will assume that the codebook $\mathcal{X}$ satisfies the per-codeword power constraint
\begin{equation} \label{eq:power_constraint}
 \frac{\|\p{x}\|^2}{n} \leq 1, \text{ for all } \p{x} \in \mathcal{X}.
\end{equation} 
The average number of channel uses to transmit each realization of $(w_1,\dots,w_K)$ is $\kappa=\frac{n}{\ell}$.
The resulting rate of transmission of each of the $K$ messages is
\begin{equation*}
 \frac{\log \left( p^\ell \right)}{n} = \frac{\ell}{n} \log p \text{ b/dim}.
\end{equation*} 
The sum rate of all the messages is $\frac{K\ell}{n}\log p$~b/dim.

The side information at $(\p{S},\sig^2)$ over a block of $\ell$ realizations of the $K$ message symbols is of the form $\sum_{k=1}^{K}s_{m,k}\p{w}_k=\p{u}_m$, $m=1,\dots,M$, where $M=\rank(\p{S})$ and $\p{u}_m \in \Fp^{\ell}$. This side information allows the receiver to conclude that the transmitted codeword must belong to the following subcode $\mathcal{X} \! \left( \p{S},\p{u}_1,\dots,\p{u}_M \right)$ of $\mathcal{X}$,
\begin{align} \label{eq:subcode_expurgated}
\Big\{ \rho(\p{w}_1,&\dots,\p{w}_K) ~ \Big\vert ~ \p{w}_1,\dots,\p{w}_K \in \Fp^{\ell}, \nonumber \\  &\sum_{k=1}^{K} s_{m,k}\p{w}_k=\p{u}_m \text{ for } m=1,\dots,M \Big\}.
\end{align}
The optimal decoder at $(\p{S},\sig^2)$ decodes the channel output vector to the nearest codeword $\p{\hat{x}}$ of this subcode, and the error probability at this receiver is the probability that the estimated message tuple $\rho^{-1}(\p{\hat{x}})$ is not equal to the transmit message $(\p{w}_1,\dots,\p{w}_K)$.
In order to achieve the optimal performance at a given receiver $(\p{S},\sig^2)$, we thus require that the expurgated code $\mathcal{X}(\p{S},\p{u}_1,\dots,\p{u}_M)$ be a good channel code for the point-to-point AWGN channel.
In the { multicast channel} that consists of multiple receivers, the side information matrix $\p{S}$ can vary from one receiver to the next, and hence, the expurgated codes can be different at each receiver, see Fig.~\ref{fig:expurg}. 
Hence, a capacity-achieving channel code $\mathcal{X}$ is such that the resulting expurgated code at every receiver is a good channel code for the AWGN channel.

{
\subsection*{Problem Statement}

\subsubsection*{Problem Setup}
Consider a common message Gaussian broadcast channel with single transmitter and $N$ receivers. The transmitter desires to multicast $K$ independent messages from a prime field $\Fb_p$ subject to the unit power constraint~\eqref{eq:power_constraint} on the transmit codeword.
Each of the $N$ receivers $(\p{S}_1,\sig_1^2),\dots,(\p{S}_N,\sig_N^2)$ has coded side information corresponding to the side information matrix $\p{S}_i \in \Fb_p^{M_i \times K}$, $i=1,\dots,N$, and experiences an additive white Gaussian noise of variance $\sig_i^2$, $i=1,\dots,N$. Without loss of generality, we assume that each of the side information matrices $\p{S}_i$ has linearly independent rows, i.e., $\rank(\p{S}_i)=M_i$. 
Using the information-theoretic arguments of~\cite{Tun_IEEE_IT_06}, which is based on the average performance of an ensemble of randomly generated codebooks, it can be shown that the (symmetric) capacity of this multicast channel is
\begin{equation} \label{eq:capacity}
 C = \min_{i \in \{1,\dots,N\}} \frac{1}{(K-M_i)}\cdot{\frac{1}{2}\log_2\left( 1 + \frac{1}{\sig_i^2} \right)}.
\end{equation} 
The proof of this result is similar to the proof of Theorem~6 of~\cite{Tun_IEEE_IT_06} which considers a {\em discrete memoryless} common message broadcast channel where the side information at each receiver is, in general, a random variable jointly distributed with the source messages $(w_1,\dots,w_K)$.
A sketch of the proof that $C$ is the capacity for the Gaussian multicast channel with coded side information at the receivers is given in the appendix. 

\subsubsection*{Problem Statement}
Let $\epsilon,\zeta>0$ be fixed positive real numbers and let $R \leq C -\epsilon$. 
We seek to determine whether there exists a lattice code for the multicast channel with coded side information at the receivers that transmits each of the $K$ messages with rate at least $(R -\epsilon)$ such that the probability of decoding error at each of the $N$ receivers is at the most $\zeta$.

In this paper we answer the above stated problem in the affirmative under the assumption that the prime field $\Fb_p$ is sufficiently large. In particular, we prove the existence of a lattice code with the said properties when the prime $p$ satisfies the inequality
$\textstyle p \geq \max\left\{ 2^{2KR} , (2^{\frac{\epsilon}{4}}-1)^{-1} 2^{-R}  \right\}$.
Unlike the capacity~\eqref{eq:capacity} which holds for any value of $p$, our result on the optimality of lattice codes requires that $p$ vary with the tolerance $\epsilon$.
The larger the gap to capacity $\epsilon$, the smaller is the size requirement on the prime field $\Fb_p$.
}%

\subsection{Lattice Preliminaries} \label{sec:lattice_prelim}

We now briefly recall the necessary properties of lattices and lattice codes, and establish our notation and terminology. The material presented in this section consists of standard ingredients used in the literature, and is mainly based on~\cite{CoS_Springer_99,For_JSAC_89,Zam_Cambridge_14,ErZ_IEEE_IT_04}.

\subsubsection{Lattices and Lattice Codes} \label{sec:prelim:lattices}

Throughout this manuscript we consider $n$-dimensional lattices $\Lambda$ with full-rank generator matrix. The closest vector lattice quantizer corresponding to $\Lambda$ is denoted by the function $Q_{\Lambda}: \Rb^n \to \Lambda$, and the volume of its (fundamental) Voronoi region $\vor(\La)=Q_{\Lambda}^{-1}(\pmb{0})$ is denoted by $\vol(\La)$. 
For any $\p{\lambda} \in \La$, $\p{\lambda} + \vor(\La)$ is the set of all points in $\Rb^n$ that are mapped to $\p{\lambda}$ under $Q_{\La}$, and it has the same volume $\vol(\La)$ as $\vor(\La)$. For any two distinct lattice points $\p{\lambda}_1 \neq \p{\lambda}_2$, the sets $\p{\lambda}_1 + \vor(\La)$ and $\p{\lambda}_2 + \vor(\La)$ are disjoint.
The \emph{modulo}-$\La$ operation, defined as $[\pmb{x}] \mmod \La=\pmb{x}-Q_{\La}(\pmb{x})$, satisfies the following properties for all $\p{x},\p{x}_1,\p{x}_2 \in \Rb^n$
\begin{align}
[\p{x}] \mmod \La &\in \vor(\La), \nonumber \\
[\p{x}_1 + \p{x}_2] \mmod \La &= \big[  \, [\p{x}_1] \mmod \La + \p{x}_2 \big] \mmod \La, \text{ and} \label{eq:mod_is_distributive} \\
[\p{x}] \mmod \La &= \p{0} \text{ if and only if } \p{x} \in \La. \label{eq:when_is_mod_zero}
\end{align} 
We will denote the $n$-dimensional ball of radius $r$ with center $\p{s} \in \Rb^n$ as $\ball(\p{s},r)$, i.e.
$\ball(\p{s},r) = \left\{ \p{x} \in \Rb^n \, \vert \, \|\p{x}-\p{s}\| \leq r \right\}$,
and the volume of a unit-radius ball in $n$ dimensions by $V_n$. It follows that the volume of $\ball(\p{s},r)$ equals $V_n r^n$.
The \emph{covering radius} of the lattice $\La$ is denoted by $\rcov(\La)$
and the \emph{effective radius} of $\La$ by $\reff(\La)$. 
We recall that $\reff(\La) \leq \rcov(\La)$ and %% is related to $\vol(\La)$ as
\begin{equation} \label{eq:reff}
 \reff(\La) = \left( \frac{\vol(\La)}{V_n} \right)^{\frac{1}{n}}.
\end{equation} 
Rogers~\cite{Rog_Math_59} showed that for every dimension $n$ there exists a lattice $\La$ such that
\begin{equation} \label{eq:rogers}
 \frac{\rcov(\La)}{\reff(\La)} \leq \left( c\,n\,({\rm ln}\,n)^{\frac{1}{2}\log 2\pi e} \right)^{\frac{1}{n}},
\end{equation} 
where $c$ is a constant. Note that the right hand side of the above inequality converges to $1$ as $n \to \infty$.
A sequence of lattices of increasing dimension $n$ is said to be \emph{Rogers-good} if $\frac{\rcov}{\reff} \to 1$. Rogers' result~\eqref{eq:rogers} shows that such a sequence exists (see also~\cite{ELZ_IT_05}).

Let $\Lc \subset \La$ be a pair of nested lattices and $\p{d} \in \vor(\Lc)$ be a fixed vector.
A \emph{(nested) lattice code} or a \emph{Voronoi code} $(\La - \pmb{d})/\Lc$ is the set $(\La-\p{d}) \mmod \Lc$ obtained by applying the $\mmod \Lc$ operation on the points of the lattice translate $\La-\p{d}$.
The code consists of all the points in $\La-\p{d}$ that lie within the Voronoi region of $\Lc$, i.e., $(\La - \pmb{d}) \cap \vor(\Lc)$. 
The lattice $\Lc$ is called the \emph{coarse lattice} or the \emph{shaping lattice}, $\La$ is called the \emph{fine lattice} or the \emph{coding lattice}, and $\p{d}$ is the \emph{dither} vector. The cardinality of this code is $|(\La-\p{d})/\Lc|=|\La/\Lc|=\frac{\vol(\Lc)}{\vol(\La)}$, and every codeword point $\pmb{x} \in (\La-\p{d})/\Lc$ satisfies $\|\pmb{x}\| \leq \rcov(\Lc)$.
Note that $\La/\Lc$ is a lattice  code with zero dither.

\subsubsection{Lattice Codes from Linear Codes over a Finite Field} \label{sec:prelim:lattices_from_codes}

In this subsection we briefly describe the method proposed in~\cite{NaG_IT_11} to construct a pair $\Lc \subset \La$ of nested lattices, and recall its relevant properties.
This construction uses a coarse lattice $\Lc$ and a linear code $\mathscr{C}$ to generate a fine lattice $\La$ such that $|\La/\Lc|=|\mathscr{C}|$.

Let $g(\cdot)$ denote the natural map that embeds $\Fp=\{0,1,\dots,p-1\}$ into $\Zb$. When applied to vectors, $g(\cdot)$ acts independently on each component of a vector.
Let $\mathscr{C} \subset \Fp^{n}$ be a linear code of rank $L$, $1 \leq L \leq n$,
\begin{equation*}
 \mathscr{C} = \left\{ \p{Gw} \, \vert \, \p{w} \in \Fp^{L}  \right\},
\end{equation*} 
where $\p{G}$ is the $n \times L$ generator matrix with full column rank, and $\p{w}$ is the message encoded to $\mathscr{C}$.
The set $g(\mathscr{C}) + p\Zb^n$ obtained by tiling copies of $g(\mathscr{C})$ at every vector of $p\Zb^n$ is a lattice in $\Rb^n$ and is known as the \emph{Construction~A} lattice of the linear code $\mathscr{C}$~\cite{CoS_Springer_99}. 
Note that the number of points in $g(\mathscr{C})+p\Zb^n$ contained in the Voronoi region of the lattice $p\Zb^n$ is $|\mathscr{C}|=p^{L}$.
We obtain $\La$ by scaling down the Construction~A lattice by $p^{-1}$ and transforming it by the generator matrix $\Bc$ of $\Lc$
\begin{align*}
\La &= \Bc p^{-1} \left( g(\mathscr{C}) + p\Zb^n \right) = \Bc p^{-1} g(\mathscr{C}) + \Bc \Zb^n \\ &= \Bc p^{-1} g(\mathscr{C}) + \Lc.
\end{align*} 
Since $\mathscr{C}$ contains the all zero codeword, it follows that $\La \supset \Bc p^{-1} g(\p{0}) + \Lc = \Lc$. 
We observe that applying the transformation $\Bc p^{-1}$ to the lattice $p\Zb^n$ (instead of the lattice $g(\mathscr{C})+p\Zb^n$) generates $\Lc$ (instead of $\La$).
Hence, $\La/\Lc$ has the same algebraic structure as that of $(g(\mathscr{C})+p\Zb^n)/p\Zb^n$, which in turn, is equivalent to the linear code $\mathscr{C}$. In particular,
\begin{equation} \label{eq:size_of_La_over_Lc}
|\La/\Lc|=|\mathscr{C}|=p^{L}.
\end{equation} 
The following lemma provides an explicit bijection between the message vectors $\p{w} \in \Fp^L$ encoded by $\mathscr{C}$ and the points in the lattice code $\La/\Lc$.
This result, which is originally from~\cite[Lemma~5]{NaG_IT_11}, is proved below for completeness.

\begin{lemma} \label{lem:w_to_t}
The map $\p{w} \to \left[ \Bc \, p^{-1} g\left( \p{Gw} \right) \right] \mmod \Lc$ is a bijection between $\Fp^{L}$ and $\La/\Lc$.
\end{lemma}
\begin{IEEEproof}
From~\eqref{eq:size_of_La_over_Lc}, we know that $|\La/\Lc|=|\Fp^{L}|=p^{L}$. Hence, it only remains to show that no two distinct messages $\p{w}_A$ and $\p{w}_B$ are mapped to the same point in $\La/\Lc$. Assuming the contrary, we have $\left[\Bc p^{-1} g(\p{Gw}_A)\right] \mmod \Lc = \left[\Bc p^{-1} g(\p{Gw}_B)\right] \mmod \Lc$. Using~\eqref{eq:mod_is_distributive} and~\eqref{eq:when_is_mod_zero}, we obtain
\begin{equation*}
\Bc p^{-1} g(\p{Gw}_A) - \Bc p^{-1} g(\p{Gw}_B) \in \Lc. 
\end{equation*} 
Multiplying both sides by $p\Bc^{-1}$, we obtain $g(\p{Gw}_A) - g(\p{Gw}_B) \in p\Zb^n$. Reducing this result modulo-$p$, we have $\p{Gw}_A-\p{Gw}_B=\p{0}$ over $\Fp$. Since this implies $\p{G}(\p{w}_A-\p{w}_B)=\p{0}$ over $\Fp$ while $\p{w}_A-\p{w}_B\neq \p{0}$ and $\p{G}$ has full column rank, we have arrived at a contradiction.
\end{IEEEproof}

In order to prove capacity achievability, we will rely on random coding arguments to show the existence of a good choice of $\p{G}$. As in~\cite{NaG_IT_11}, we will assume that $\p{G}$ is a random matrix chosen with uniform probability distribution on $\Fp^{n \times L}$. The following result is useful in upper bounding the decoding error probability over the ensemble of random codes.

\begin{lemma}[\cite{ELZ_IT_05,NaG_IT_11,KrP_Good_Lattices_07}] \label{lem:t_is_uniform}
Let $\p{w} \in \Fp^{L} \backslash \{\p{0}\}$ be a given non-zero vector, and let $\p{G}$ be uniformly distributed in $\Fp^{n \times L}$. Then $\left[\Bc p^{-1} g(\p{Gw}) \right] \! \mmod \Lc$ is uniformly distributed over $\left(p^{-1}\Lc\right) \cap \vor(\Lc)$, i.e., over the lattice code $p^{-1}\Lc/\Lc$.   
\end{lemma} 

\section{Lattice Codes for the { Common Message Gaussian Broadcast Channel} with Coded Side Information} \label{sec:3}

We will assume that the number of messages $K$ and a design rate $R$ are given, and show that there exist good lattice codes of sufficiently large dimension $n$ that encode $K$ messages over an appropriately chosen prime field $\Fp$ at rates close to $R$~b/dim. 
%% R1-EDIT %%
In order to rigorously state the main result, we consider a fixed non-zero tolerance $\epsilon>0$ that determines the gap to capacity. %% and the probability of decoding error.

\begin{theorem}[Main theorem] \label{thm:main}
Let the number of messages $K$, design rate $R$ and tolerance $\epsilon>0$ be given. 
For every sufficiently large prime integer $p$,
there exists a sequence of lattice codes of increasing dimension $n$ %% $(\La-\p{d})/\Lc$ %and
that encode $K$ message vectors over $\Fp$ %into $(\La-\p{d})/\Lc$,
such that the rate of transmission of each message is at least $(R-\epsilon)$~b/dim and the probability of error at a receiver $(\p{S},\sig^2)$ decays exponentially in $n$ if
\begin{align} 
\frac{1}{2} \log \left(1+\frac{1}{\sig^2}\right) \geq (R + \epsilon) \, (K - \rank(\p{S})). \label{eq:snr_rate_condition}
\end{align} 
\end{theorem}

To prove Theorem~\ref{thm:main}, we utilize the lattice code ensemble introduced in~\cite{NaG_IT_11}; see Section~\ref{sec:prelim:lattices_from_codes} of this paper. 
A Rogers' good lattice is chosen as the coarse lattice $\Lc$. The fine lattice $\La$ is obtained from the generator matrix $\Bc$ of the coarse lattice $\Lc$ and a linear code $\mathscr{C}$ over a large enough prime field $\Fb_p$ using the construction described in Section~\ref{sec:prelim:lattices_from_codes}. 

{
The multicast channel considered in Theorem~\ref{thm:main} reduces to the traditional single-user AWGN channel if the number of messages $K=1$, and the multicast channel consists of one receiver with an empty side information matrix $\p{S}$, i.e., $\rank(\p{S})=0$. Hence, Theorem~\ref{thm:main} provides an alternative proof of the existence of lattice codes that achieve the capacity of the single-user AWGN channel, and we have the following corollary. 

\begin{corollary} \label{cor:awgn_channel}
Consider a single user AWGN channel where the input is subject to unit power constraint and the noise variance at the receiver is $\sigma^2$. Let $\epsilon>0$ be any constant and let 
\begin{equation*}
R \leq  \frac{1}{2} \log \left(1+\frac{1}{\sig^2}\right) - \epsilon.
\end{equation*} 
For every sufficiently large prime integer $p$, there exists a sequence of lattice codes of increasing dimension $n$ constructed from linear codes over $\Fb_p$ (as described in Section~\ref{sec:prelim:lattices_from_codes}) such that the rate of each lattice code is at least $R-\epsilon$ and the probability of decoding error at the receiver decays exponentially in $n$.
\end{corollary} 

The relation of Corollary~\ref{cor:awgn_channel} to existing results on the optimality of Construction~A based lattice codes in single-user AWGN channel is described in detail in Section~\ref{sec:corollaries:awgn}.
}%

In the rest of this section we describe the construction of random lattice codes, and the encoding and decoding operations used to prove Theorem~\ref{thm:main}. We provide the proof of the Theorem~\ref{thm:main} in Section~\ref{sec:proof_of_main}.

\begin{figure*}[!t]
\centering
\includegraphics[width=5in]{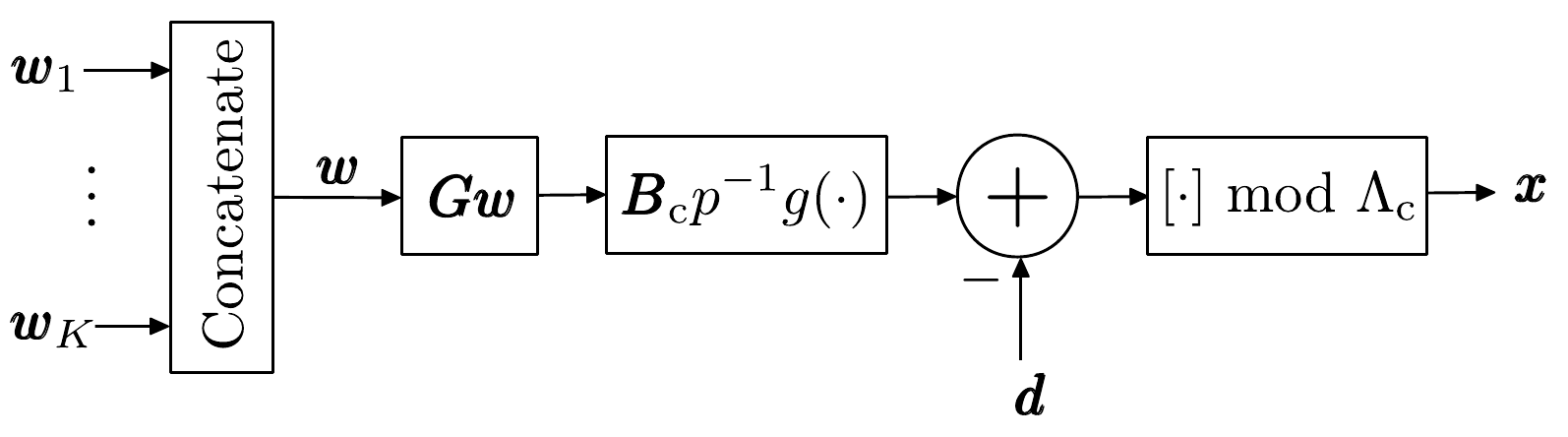}
\caption{The encoding operation at the transmitter that maps the messages $\p{w}_1,\dots,\p{w}_K$ to a point in $(\La-\p{d})/\Lc$.}
\label{fig:encoder}
\end{figure*} 

\subsection{Random lattice code ensemble} \label{sec:sub:choice_of_parameters}

\subsubsection{Prime $p$}

{
Given the design rate $R$, number of messages $K$ and tolerance $\epsilon>0$, we 
% choose $p$ as any prime integer satisfying
require $p$ to satisfy the constraint
\begin{equation}~\label{eq:prime_constraint}
 p \geq \max\left\{ 2^{2KR} , (2^{\frac{\epsilon}{4}}-1)^{-1} 2^{-R}  \right\}.
\end{equation} }
The coding schemes of this paper are based on Construction~A lattices which are obtained by lifting linear codes over $\Fb_p$ to the Euclidean space $\Rb^n$. The generator matrices of these $\Fb_p$-linear codes are constructed randomly, and the first constraint in~\eqref{eq:prime_constraint}, viz. 
$p \geq 2^{2KR}$,
will allow us to show that these randomly constructed generator matrices are full-ranked with probability close to $1$.

The proof of Theorem~\ref{thm:main} given in Section~\ref{sec:proof_of_main} involves the derivation of an upper bound on the probability of decoding error averaged over an ensemble of lattice codes derived from Construction~A. 
We will use the inequality $p \geq (2^{\frac{\epsilon}{4}}-1)^{-1} 2^{-R}$ from~\eqref{eq:prime_constraint} to show that this upper bound is exponentially small in dimension $n$. Note that this inequality implies
\begin{equation*}
 p \geq (2^{\frac{\epsilon}{4}}-1)^{-1} 2^{-(R+\epsilon)(K-M)}
\end{equation*} 
for any integer $M$ satisfying $0 \leq M \leq K-1$. Rearranging the terms in the above inequality we obtain
\begin{equation} \label{eq:prime_constraint_2}
\frac{1}{p\,2^{(R+\epsilon)(K-M)}} + 1 \leq 2^{\frac{\epsilon}{4}}.
\end{equation} 

\subsubsection{Message length $\ell$}

Once $p$ is fixed, we choose $\ell$ as the largest integer that satisfies
\begin{equation} \label{eq:ell_over_n_rate}
\frac{\ell}{n} \log p \leq R.
\end{equation} 
The left-hand side in the above inequality is the actual rate at which the lattice code encodes each message, while $R$ is the design rate. The difference between the two is at the most
\begin{equation*}
 \frac{\ell+1}{n} \log p - \frac{\ell}{n} \log p = \frac{\log p}{n} %% R1-EDIT \leq \frac{\log\left( n^\beta \right)}{n} = \frac{\beta \, \log n}{n},
\end{equation*} 
which converges to $0$ as $n \to \infty$. 
{%%
It follows that the code rate $\frac{\ell}{n} \, \log_2 p$ tends to the design rate $R$ as $n \to \infty$, and hence, $\frac{\ell}{n} \, \log_2 p \geq R-\epsilon$ for all sufficiently large $n$.
}%

\subsubsection{Coarse Lattice $\Lc$}

From~\eqref{eq:rogers} in Section~\ref{sec:prelim:lattices}, we know that for a given $\epsilon>0$ and for all sufficiently large $n$, there exists an $n$-dimensional lattice $\Lc$ such that 
\begin{equation*}
\frac{\rcov(\Lc)}{\reff(\Lc)} \leq 2^{\frac{\epsilon}{4}}.
\end{equation*} 
We will choose such a Rogers-good lattice as $\Lc$, and scale it so that 
\begin{equation*}
\rcov(\Lc)=\sqrt{n}.
\end{equation*} 
It follows that $\reff(\Lc) \geq 2^{-\frac{\epsilon}{4}} \, \rcov(\Lc) = 2^{-\frac{\epsilon}{4}} \sqrt{n}$. Using the definition of the effective radius~\eqref{eq:reff}, we arrive at the following lower bound on the volume of the Voronoi region of $\Lc$
\begin{equation} \label{eq:vol_Lc_bound}
\vol(\Lc) = V_n \, \reff^n(\Lc) \geq V_n \,n^{\frac{n}{2}} \, 2^{-\frac{n\epsilon}{4}}.
\end{equation} 

\subsubsection{Fine Lattice $\La$} 

The fine lattice is obtained by the construction of~\cite{NaG_IT_11} described in~Section~\ref{sec:prelim:lattices_from_codes}. The length of the linear code $\mathscr{C}$ is $n$, and its rank $L=K\ell$ is the number of message symbols to be encoded by the lattice code. 
Note that this requires that $K\ell < n$ be true.
{%%
Using~\eqref{eq:ell_over_n_rate} and the property $p \geq 2^{2KR}$, we have
\begin{equation} \label{eq:K_ell_bound}
 K\ell \leq \frac{nKR}{\log p} \leq \frac{nKR}{2KR} = \frac{n}{2} %% R1-EDIT \frac{nKR}{\log\left( \frac{n^\beta}{2} \right)} = \frac{nR}{\beta\log n - 1}.
\end{equation}
%% R1-EDIT %%
which ensures that $K\ell < n$.
}
%% END %%
If $\p{G} \in \Fp^{n \times K\ell}$ is the generator matrix of $\mathscr{C}$, then $\La = \Bc p^{-1} g(\mathscr{C}) + \Lc$. We will choose $\p{G}$ uniformly random over the set of all $n \times K\ell$ matrices of $\Fb_p$, resulting in a random ensemble of fine lattices $\La$.

\subsubsection{Dither vector $\p{d}$}

We will rely on random coding arguments to prove the existence of a translate $\p{d}$ such that the code $(\La -\p{d})/\Lc$ performs close to capacity. We will assume that $\p{d}$ is distributed uniformly in $\vor(\Lc)$ and is chosen independently of $\p{G}$. This random dither $\p{d}$ is usually viewed as a common randomness available at the transmitter and the receivers~\cite{ErZ_IEEE_IT_04}. Note that $\|\p{d}\| \leq \rcov(\Lc) =\sqrt{n}$.

\subsection{Encoding} \label{sec:sub:encoding}

We will now describe the encoding operation $\rho$ at the transmitter that maps the message vectors $(\p{w}_1,\dots,\p{w}_K) \in \Fp^\ell \times \cdots \times \Fp^\ell$ to a codeword \mbox{$\p{x} \in (\La-\p{d})/\Lc$}. The encoder first concatenates the $K$ messages into the vector $\p{w}=\begin{pmatrix} \p{w}_1^\tr,\cdots,\p{w}_K^\tr\end{pmatrix}^\tr$, encodes $\p{w}$ to a codeword in the linear code $\mathscr{C}$, and maps it to a point $\p{t} \in \Rb^n$ using Construction~A as follows
\begin{equation} \label{eq:w_to_t}
\p{t} = \left[ \Bc \, p^{-1} g\left( \p{Gw} \right) \right] \mmod \Lc.
\end{equation} 
From the discussion in Section~\ref{sec:sub:choice_of_parameters}, we know that $\Bc p^{-1} g(\p{Gw}) \in \La$, and hence, $\p{t} \in \La/\Lc$. Finally, the transmit codeword $\p{x}$ is generated by dithering $\p{t}$,
\begin{equation} \label{eq:w_to_x}
 \p{x} = \left[ \p{t} - \p{d} \right] \mmod \Lc = \left[ \Bc \, p^{-1} g\left( \p{Gw} \right) - \p{d} \right] \mmod \Lc.
\end{equation} 
This sequence of operations is illustrated in Fig.~\ref{fig:encoder}.
Note that since $\rcov(\Lc)=\sqrt{n}$, each codeword $\p{x}$ satisfies $\|\p{x}\| \leq \rcov(\Lc)=\sqrt{n}$, and hence, the power constraint 
${\|\p{x}\|^2}/{n} \leq 1$.
It is straightforward to show that the dithering operation~\eqref{eq:w_to_x} is a one-to-one correspondence between $\p{t} \in \La/\Lc$ and $\p{x} \in (\La-\p{d})/\Lc$. 
Further, from Lemma~\ref{lem:w_to_t} we know that~\eqref{eq:w_to_t} is a bijection between the message space $\Fp^{K\ell}$ and the undithered codewords $\La/\Lc$ if $\p{G}$ is full rank.
Hence, to ensure that no two messages are mapped to the same codeword, we only require that the random matrix $\p{G}$ be full rank.
It can be shown that (see~\cite{ELZ_IT_05})
\begin{equation*}
\Pp \left( \rank(\p{G}) < K\ell\right) \leq p^{-(n-K\ell)}.
\end{equation*} 
We will only require a relaxation based on the above inequality. From~\eqref{eq:K_ell_bound}, we have $K\ell \leq \frac{n}{2}$. Similarly, since $p$ is a prime integer, we have $p \geq 2$, and hence,
\begin{equation} \label{eq:prob_rank_def}
 \Pp \left( \rank(\p{G}) < K\ell\right) \leq 2^{-\left(n-\frac{n}{2}\right)} = 2^{-\frac{n}{2}}.
\end{equation} 

\subsection{Decoding}

% We now describe the decoding operation at a generic receiver $(\p{S},\sig^2)$ which utilizes the approach of MMSE scaling followed by decoding to an infinite lattice constellation~\cite{ErZ_IEEE_IT_04}. 
The receiver employs a two stage decoder: in the first stage the receiver identifies the subcode of $(\La-\p{d})/\Lc$ corresponding to the available side information, and in the second stage it decodes the channel output to a point in this subcode.
%% The decoding methodology also applies to the case of a receiver with no side information, i.e., when $\p{S}$ is the empty matrix. 

\subsubsection{Using Side Information to Expurgate Codewords}

% We first describe how receiver side information is used to reduce the number of candidate codewords at the decoder.
The side information at $(\p{S},\sig^2)$ over a block of $\ell$ realizations of the $K$ messages is of the form
\begin{equation} \label{eq:side_inf_vector_form:1}
\sum_{k=1}^{K} s_{m,k} \p{w}_k = \p{u}_m, ~~~\, m=1,\dots,M.
\end{equation} 
% where $\p{u}_m \in \Fp^{\ell}$, $\p{S}=[s_{k,m}] \in \Fp^{K \times M}$ and $M=\rank(\p{S})$. 
The receiver desires to identify the set of all possible values of the message vector $\p{w}=\begin{pmatrix} \p{w}_1^\tr,\cdots,\p{w}_K^\tr \end{pmatrix}^\tr$ that satisfy~\eqref{eq:side_inf_vector_form:1}. Using the notation $\p{u}=\begin{pmatrix} \p{u}_1^\tr,\cdots,\p{u}_M^\tr\end{pmatrix}{}^\tr \in \Fp^{M\ell}$, the side information~\eqref{eq:side_inf_vector_form:1} can be rewritten compactly in terms of $\p{w}$ and $\p{u}$ as
\begin{equation} \label{eq:side_inf_vector_form:2}
 \left( \p{S} \otimes \p{I}_{\ell} \right)\p{w} = \p{u},
\end{equation} 
where $\otimes$ denotes the Kronecker product of matrices and $\p{I}_\ell$ is the $\ell \times \ell$ identity matrix over $\Fp$. 
Observe that~\eqref{eq:side_inf_vector_form:2} is an under-determined system of linear equations, and the set of solutions is a coset of the null space of $\p{S} \otimes \p{I}_{\ell}$. Let $\p{A_S} \in \Fp^{K\ell \times (K-M)\ell}$ be a rank $(K-M)\ell$ matrix such that $(\p{S} \otimes\p{I}_{\ell})\p{A_S}=\p{0}$, i.e., the columns of $\pmb{A_S}$ form a basis of the null space of $\p{S}\otimes\p{I}_{\ell}$. Then the set of all solutions to~\eqref{eq:side_inf_vector_form:2} is
\begin{equation} \label{eq:set_of_solutions_w}
\p{v} + \left\{ \p{A_S \tilde{w}} \, \vert \, \p{\tilde{w}} \in \Fp^{(K-M)\ell} \right\},
\end{equation} 
where $\p{v}$ is the coset leader.
From~\eqref{eq:w_to_t}, we conclude that the undithered codeword must be of the form 
\begin{equation} \label{eq:set_T:1}
\p{t} = \left[ \Bc p^{-1} g\left( \p{Gv} + \p{GA_S\tilde{w}} \right) \right] \! \mmod \Lc,~  \p{\tilde{w}} \in \Fp^{(K-M)\ell}. 
\end{equation}
We will now use the property of $g(\cdot)$ that for any $\p{a},\p{b} \in \Fp^n$, 
\begin{equation*}
g(\p{a}+\p{b})=g(\p{a})+g(\p{b}) \mmod p. 
\end{equation*} 
Therefore, $g(\p{Gv} + \p{GA_S\tilde{w}})=g(\p{Gv}) + g(\p{GA_S\tilde{w}}) + p\p{c}$ for some $\p{c} \in \Zb^n$. 
Using this in~\eqref{eq:set_T:1}, we obtain
\begin{align}
\p{t}=&  \left[ \, \Bc p^{-1} g\left( \p{Gv} \right) + \Bc p^{-1} g\left(\p{GA_S\tilde{w}} \right) + \Bc\p{c} \, \right] \! \mmod \Lc   \nonumber \\
=& \Big[\, \Bc p^{-1} g\left( \p{Gv} \right) + \left[\Bc p^{-1} g\left(\p{GA_S\tilde{w}} \right)\right] \! \mmod \Lc \nonumber \\  &~~~~~~~~~~~~~~~~~~~~~~~~~~~~~~~~~~+ \left[\, \Bc\p{c} \, \right]\! \! \mmod \Lc  \, \Big] \mmod \Lc  \nonumber \\
=& \left[\, \Bc p^{-1} g\left( \p{Gv} \right) + \left[\Bc p^{-1} g\left(\p{GA_S\tilde{w}} \right)\right] \! \mmod \Lc \, \right] \mmod \Lc,  \label{eq:subcode_of_valid_codewords}
\end{align} 
where we have used~\eqref{eq:mod_is_distributive},~\eqref{eq:when_is_mod_zero} and the fact that $\Bc\p{c} \in \Lc$. 
Since the receiver knows $\p{v}$, the component of $\p{t}$ unavailable from the side information is
\begin{equation} \label{eq:tilde_w_to_tilde_t}
 \p{\tilde{t}} = \left[\Bc p^{-1} g\left(\p{GA_S\tilde{w}} \right)\right] \! \mmod \Lc.
\end{equation} 
Let $\CS \subset \Fp^n$ be the subcode of $\mathscr{C}$ with generator matrix $\p{GA_S}$, and $\LS$ be the lattice obtained by applying Construction~A to $\CS$ and transforming it by $\Bc p^{-1}$, i.e.,
\begin{equation*}
 \LS = \Bc p^{-1} g(\CS) + \Lc.
\end{equation*}  
Using $\p{GA_S}$ instead of $\p{G}$ in Lemma~\ref{lem:w_to_t}, we see that $\p{\tilde{t}} \in \LS/\Lc$ and that~\eqref{eq:tilde_w_to_tilde_t} is a one-to-one correspondence between $\p{\tilde{w}} \in \Fp^{(K-M)\ell}$ and $\p{\tilde{t}} \in \LS/\Lc$ as long as $\p{GA_S}$ is full rank.
Together with~\eqref{eq:w_to_x},~\eqref{eq:subcode_of_valid_codewords}, and~\eqref{eq:tilde_w_to_tilde_t}, we conclude that the transmit vector $\p{x}$ belongs to the following lattice subcode of $(\La-\p{d})/\Lc$,
\begin{align} \label{eq:x_side_inf}
 &\left( \LS + \Bc p^{-1} g\left( \p{Gv} \right) - \p{d} \right)/\Lc = \nonumber \\ 
 &~~~~~~~~~~\left\{ \left[\, \p{\tilde{t}} + \Bc p^{-1} g(\p{Gv}) -\p{d} \,\right] \! \mmod \Lc~\big\vert~ \p{\tilde{t}} \in \LS/\Lc \right\}. 
\end{align}  
The decoding problem at the second stage is to estimate $\p{\tilde{t}}$, or equivalently $\p{\tilde{w}}$, from the channel output.

\subsubsection{MMSE Scaling and Lattice Decoding}

\begin{figure*}[!t]
\centering
\includegraphics[width=5.5in]{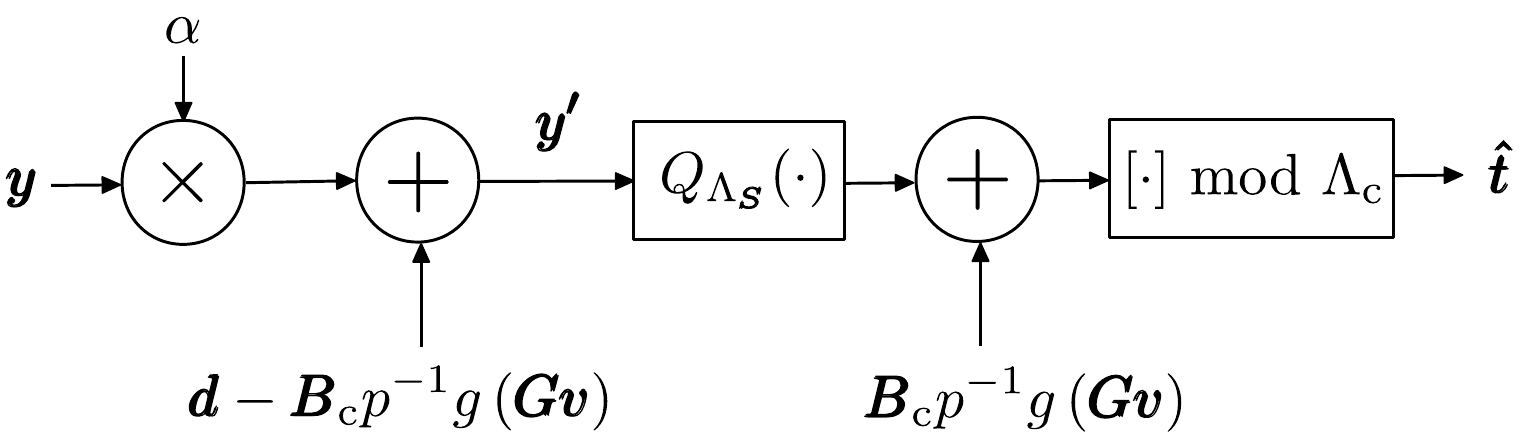}
\caption{The decoding operation at the receiver $(\p{S},\sig^2)$. The vector $\p{v}$ and the lattice $\LS$ are determined using the side information available at the receiver.}
\label{fig:decoder}
\end{figure*} 

Let the channel output at the receiver $(\p{S},\sig^2)$ be $\p{y}=\p{x}+\p{n}$, where $\p{n}$ is a Gaussian vector with zero mean and variance $\sig^2$ per dimension. The received vector is scaled by the coefficient $\al$, resulting in
\begin{equation} \label{eq:MMSE_scaling}
\al\p{y} = \al\p{x} + \al\p{n} = \p{x} + \al\p{n} - (1-\al)\p{x}.
\end{equation} 
This MMSE pre-processing improves the effective signal-to-noise ratio of the system beyond the channel signal-to-noise ratio $\frac{1}{\sig^2}$ and allows the lattice decoder to perform close to capacity~\cite{ErZ_IEEE_IT_04,For_Allerton_03}. Let
\begin{equation*}
\p{z} = \al\p{n} - (1-\al)\p{x} 
\end{equation*} 
be the effective noise term in~\eqref{eq:MMSE_scaling}. Using the facts that $\p{x}$ and $\p{n}$ are independent, $\|\p{x}\| \leq \sqrt{n}$, and $\p{n}$ has zero mean, we have
\begin{align*}
\Eb \, \|\p{z}\|^2 &= (1-\al)^2 \, \Eb \, \|\p{x}\|^2 + \al^2 \Eb \, \|\p{n}\|^2 - 2 \al(1-\al)\,\Eb \,\p{x}^\tr\p{n} \\
                   &\leq (1-\al)^2 \, n + \al^2 \sig^2 \,n,
\end{align*} 
where $\Eb$ is the expectation operator. The choice of $\al={1}/{(1+\sig^2)}$ minimizes this upper bound and yields
% \begin{align*}
\mbox{$\Eb \, \|\p{z}\|^2 \leq {n\,\sig^2}/{(1+\sig^2)}$},
% \end{align*} 
which is less than the Gaussian noise power $\Eb \|\p{n}\|^2 = n\sig^2$. In the rest of the paper we will assume that $\alpha={1}/{(1+\sig^2)}$ and use the notation
\begin{equation} \label{eq:sigz}
\sigz^2 = \frac{\sig^2}{1+\sig^2}.
\end{equation}
The lower bound~\eqref{eq:snr_rate_condition} on signal-to-noise ratio can be rewritten in terms of $\sigz^2$ as
\begin{equation} \label{eq:sigz_upper_bound}
 \sigz^2 \leq 2^{-2(R+\epsilon)(K-M)}.
\end{equation} 

% We will show that the probability that $\|\p{z}\|^2$ is much larger than its mean ${n\sigz^2}$ is exponentially small in $n$.

From~\eqref{eq:x_side_inf}, we know that $\p{x} = \p{\tilde{t}} + \Bc p^{-1}g(\p{Gv}) - \p{d} + \lc$ for some $\lc \in \Lc$. 
After MMSE scaling, the decoder removes the contributions of the dither $\p{d}$ and the offset $\Bc p^{-1} g(\p{Gv})$ from $\al\p{y}$ to obtain
\begin{align*}
 \p{y'} &= \al\p{y} - \Bc p^{-1} g(\p{Gv}) + \p{d} 
        = \p{\tilde{t}} + \lc + \p{z}.
\end{align*} 
The decoder proceeds by quantizing $\p{y'}$ to the lattice $\LS$ and reducing the result modulo $\Lc$. If the noise $\p{z}$ is sufficiently `small', then this sequence of operations will yield
\begin{align} \label{eq:est_t_tilde}
\left[ Q_{\LS}(\p{y'}) \right] \mmod \Lc &= \left[ Q_{\LS}(\p{\tilde{t}} + \lc + \p{z}) \right] \mmod \Lc \nonumber \\ &= \left[ \p{\tilde{t}} + \lc \right] \mmod \Lc = \p{\tilde{t}}.
\end{align}
Given $\p{\tilde{t}}$, the receiver uses~\eqref{eq:subcode_of_valid_codewords} to obtain the undithered codeword $\p{t}$, and hence the message vector $(\p{w}_1^\tr,\dots,\p{w}_K^\tr)^\tr$, as
$\p{t} = \left[\, \Bc p^{-1} g(\p{Gv}) + \p{\tilde{t}}  \,\right] \mmod \Lc$.
To conclude, the decoder obtains the estimate $\p{\hat{t}}$ of the undithered codeword $\p{t}$ from the received vector $\p{y}$ as
\begin{align*}
\p{\hat{t}} &= \Big[ \, \left[ Q_{\LS}(\p{y'}) \right] \mmod \Lc + \Bc p^{-1} g(\p{Gv})  \, \Big] \mmod \Lc \\
            &= \left[ \, Q_{\LS}(\p{y'}) + \Bc p^{-1} g(\p{Gv})  \, \right] \mmod \Lc \\
            &= \left[ \, Q_{\LS} \!\! \left(\alpha \p{y} - \Bc p^{-1} g(\p{Gv}) + \p{d}\right) + \Bc p^{-1} g(\p{Gv})  \, \right] \!\! \mmod \Lc
\end{align*} 
which shows that the $\mmod \Lc$ operation arising from~\eqref{eq:est_t_tilde} can be ignored.
The steps involved in the decoding operation are illustrated in Fig.~\ref{fig:decoder}.

Note that the effective information vector $\p{\tilde{w}}$ is not encoded in the point $\p{\tilde{t}} \in \LS$, but is encoded in the coset $\p{\tilde{t}} + \Lc$. The error event for this decoder is $Q_{\LS}(\p{y'}) \notin \p{\tilde{t}} + \Lc$, i.e., \mbox{$Q_{\LS}(\p{\tilde{t}} + \p{\lambda}_{\rm c} + \p{z}) \notin \p{\tilde{t}} + \Lc$}, which is equivalent to the event \mbox{$Q_{\LS}(\p{z}) \notin \Lc$}. Hence, a decoding error occurs if and only if $\p{z}$ is closer to a point in $\LS \backslash \Lc$ than any vector in the coarse lattice $\Lc$, i.e., if and only if
\begin{equation} \label{eq:error_event}
\mathcal{E} \!: Q_{\LS}(\p{z}) \in \LS \backslash \Lc. 
\end{equation} 

\section{Proof of Main Theorem} \label{sec:proof_of_main}

In this section we first state and prove two technical lemmas (Section~\ref{sec:technical_lemma}), use these lemmas to show that the error probability at a given fixed receiver $(\p{S},\sig^2)$ is small (Section~\ref{sec:error_at_single_receiver}), and then complete the proof of the main theorem by showing that the error probability at every receiver of the { multicast channel} is simultaneously small (Section~\ref{sec:complete_the_proof}). Finally, we state some important corollaries of the main theorem (Section~\ref{sec:corollaries}).

\subsection{Technical Lemmas} \label{sec:technical_lemma}

The first result, which is a direct generalization of~\cite[Lemma~1]{OrE_IT_16} and~\cite[Lemma~2.3]{diP_thesis_14}, gives an upper bound on the number of lattice points lying inside a ball. 

\begin{lemma} \label{lem:counting}
For any $\p{s} \in \Rb^n$, $r>0$ and any $n$-dimensional lattice $\Lc$,
\begin{equation*}
|\Lc \cap \ball(\p{s},r)| \leq \frac{V_n}{\vol(\Lc)} \left( \rcov(\Lc) + r \right)^n,
\end{equation*} 
where $V_n$ is the volume of a unit ball in $\Rb^n$.
\end{lemma}
\begin{IEEEproof}
Let $\mathcal{R}=\left( \Lc \cap \ball(\p{s},r) \right) + \vor(\Lc)$ be the set of all points in $\Rb^n$ that are mapped to one of the points in \mbox{$\Lc \cap \ball(\p{s},r)$} by the lattice quantizer $Q_{\Lc}$. Since $\mathcal{R}$ is a union of the pairwise disjoint sets $\p{\lambda} + \vor(\Lc)$, $\p{\lambda} \in \Lc \cap \ball(\p{s},r)$, and since each of these sets has volume $\vol(\Lc)$, we have 
\begin{equation} \label{eq:lem:sphere_counting:1}
\vol(\mathcal{R})=\vol(\Lc)~|\Lc \cap \ball(\p{s},r)|. 
\end{equation} 
Using the fact that \mbox{$\vor(\Lc) \subset \ball(\p{0},\rcov(\Lc))$}, we have %% the following
\begin{align*}
\mathcal{R} &= \left( \Lc \cap \ball(\p{s},r) \right) + \vor(\Lc)  
              ~\subset~ \ball(\p{s},r) + \vor(\Lc) \\
            & \subset~ \ball(\p{s},r) + \ball(\p{0},\rcov(\Lc)) 
             ~\subset~ \ball(\p{s},r+\rcov(\Lc)),
\end{align*} 
where the last step follows from triangle inequality. Consequently, we have an upper bound on the volume of $\mathcal{R}$, 
\begin{equation*}
\vol(\mathcal{R}) \leq \vol\left(\,\ball\left(\p{s},r+\rcov(\Lc)\right)\,\right)= V_n(r+\rcov(\Lc))^n. 
\end{equation*} 
Using this result with~\eqref{eq:lem:sphere_counting:1} proves the lemma.
\end{IEEEproof}

As in~\cite{Loe_IT_97,OrE_IT_16,diP_thesis_14}, we will rely on the fact that, with very high probability, the norm of the noise $\p{z}$ is not much larger than $\sqrt{n\sigz^2}$. 
The probability that the effective noise $\p{z}$ is `large' is exponentially small in $n$.  
The proof of this result is given below. 

\begin{lemma} \label{lem:noise_is_small}
Let $\p{x}$ be uniformly distributed in $\vor(\Lc)$ and $\delta>0$ be any positive number. Then
\begin{align} \label{eq:lem:z_is_not_too_large}
\Pp \! \left( \|\p{z}\|^2 > n \sigz^2 (1+\delta )\right) &\leq e^{-\frac{n \left( \delta - {\rm ln}(1+\delta) \right)}{2} } + e^{-\frac{n\sig^2\delta^2}{4}}.
\end{align} 
\end{lemma}
\begin{IEEEproof}
We will prove~\eqref{eq:lem:z_is_not_too_large} for every fixed realization of $\p{x}$ in $\vor(\Lc)$, which shows that the statement of the lemma is true for any distribution of $\p{x}$ on $\vor(\Lc)$. In the rest of the proof we will assume that $\p{x} \in \vor(\Lc)$ is an arbitrary fixed vector and $\p{n}$ is Gaussian distributed.
Using $\|\p{x}\|^2 \leq \rcov^2(\Lc)=n$, we have
\begin{align*}
 \|\p{z}\|^2 &= \|\al\p{n}-(1-\al)\p{x}\|^2 \\ 
             &=  \al^2\|\p{n}\|^2 + (1-\al)^2\|\p{x}\|^2 - 2\al(1-\al)\p{x}^\tr\p{n} \\
             &\leq \al^2\|\p{n}\|^2 + (1-\al)^2 n - 2\al(1-\al)\p{x}^\tr\p{n}.
\end{align*}
Hence, $\Pp\left( \|\p{z}\|^2 > n \sigz^2 (1+\delta) \right)$ is upper bounded by
\begin{align*}
\Pp\left( \al^2\|\p{n}\|^2 + (1-\al)^2 n - 2\al(1-\al)\p{x}^\tr\p{n} > n \sigz^2 (1+\delta) \right).
\end{align*} 
From the definition~\eqref{eq:sigz} of $\sigz^2$, we have $n\sigz^2(1+\delta)=n\al^2\sig^2(1+\delta) + n(1-\al)^2(1+\delta)$. Hence, the above upper bound corresponds to the event
\begin{align} \label{eq:lem:z_is_not_too_large:1}
 \al^2\|\p{n}\|^2 + (1-\al)^2 &n - 2\al(1-\al)\p{x}^\tr\p{n} ~> \nonumber \\ &n\al^2\sig^2(1+\delta) + n(1-\al)^2(1+\delta).
\end{align}
The event~\eqref{eq:lem:z_is_not_too_large:1} occurs \emph{only if} at least one of the following two events occurs
\begin{align}
 \mathcal{E}_A \! &: \, \al^2\|\p{n}\|^2 \, > \, n \al^2\sig^2(1+\delta), \, \text{ or } \label{eq:lem:noise_not_large:1} \\
 \mathcal{E}_B \! &: \, (1-\al)^2 n - 2\al(1-\al)\p{x}^\tr\p{n} \, > \, n(1-\al)^2(1+\delta). \nonumber
\end{align} 
Therefore, 
\begin{equation*}
\Pp(\|\p{z}\|^2 > n \sigz^2(1+\delta)) \leq \Pp(\mathcal{E}_A \cup \mathcal{E}_B) \leq \Pp(\mathcal{E}_A) + \Pp(\mathcal{E}_B).
\end{equation*} 
We will now individually upper bound $\Pp(\mathcal{E}_A)$ and $\Pp(\mathcal{E}_B)$, and thereby complete the proof.

A rearrangement of terms in~\eqref{eq:lem:noise_not_large:1} yields $\Pp(\mathcal{E}_A) = \Pp\left( \|\frac{1}{\sig} \, \p{n} \|^2 > n (1+\delta)  \right)$. This is the probability that a Gaussian vector with unit variance per dimension lies outside the sphere of squared radius $n(1+\delta)$. The following is a well known upper bound on this probability (see~\cite{Pol_IT_94})
\begin{align*}
\Pp(\mathcal{E}_A) \leq e^{-\frac{n \left( \delta - {\rm ln}(1+\delta) \right)}{2}}.
\end{align*} 
% Note that $\delta - {\rm ln}(1+\delta)>0$ for all $\delta>0$.

The event $\mathcal{E}_B$ is equivalent to $-2\al(1-\al)\p{x}^\tr\p{n} > n(1-\al)^2\delta$. Using $\al={1}/{(1+\sig^2)}$, we can show that this is same as $\p{x}^\tr\p{n} < -{n\delta\sig^2}/{2}$. Since $\p{x}^\tr\p{n}$ is a zero mean Gaussian random variable with variance $\sig^2\|\p{x}\|^2$, we have
\begin{equation*}
 \Pp(\mathcal{E}_B) = \Pp\left(\p{x}^\tr\p{n} < -\frac{n\delta\sig^2}{2}\right) = \mathcal{Q} \left( \frac{n\sig\delta}{2\|\p{x}\|} \right),
\end{equation*} 
where $\mathcal{Q}(\cdot)$ is the Gaussian tail function. Using $\|\p{x}\| \leq \rcov(\Lc) = \sqrt{n}$ and the Chernoff bound $\mathcal{Q}(y) \leq \exp(-{y^2}/{2})$, we arrive at
\begin{align*}
 \Pp(\mathcal{E}_B) \leq e^{-\frac{n\sig^2\delta^2}{4}}.
\end{align*} 
This completes the proof.
\end{IEEEproof}

\subsection{Error probability at a single receiver} \label{sec:error_at_single_receiver}

In this subsection we derive an upper bound on the decoding error probability $\Pp_{\p{S}}$ at a receiver $(\p{S},\sig^2)$ when averaged over the ensemble of lattice codes generated by choosing $\p{G}$ uniformly over $\Fp^{n \times K\ell}$ and $\p{d}$ uniformly over $\vor(\Lc)$. 

The following result from~\cite{ErZ_IEEE_IT_04}, known as the \emph{Crypto lemma}, captures an important characteristic of random dithering.
\begin{lemma}[\cite{ErZ_IEEE_IT_04}] \label{lem:crypto}
Let $\p{t} \in \vor(\Lc)$ be any random vector. If $\p{d}$ is independent of $\p{t}$ and is uniformly distributed over $\vor(\Lc)$, then $\p{x} = [\p{t} - \p{d}] \! \mmod \Lc$ is independent of $\p{t}$ and uniformly distributed over $\vor(\Lc)$.
\end{lemma}
The property that the transmit vector $\p{x}$ is statistically independent of $\p{t}$ implies that the effective noise $\p{z}=\al\p{n}-(1-\al)\p{x}$ is independent of the transmit message. 
This facilitates the error probability analysis through the observation that the error event~\eqref{eq:error_event} is statistically independent of $\p{\tilde{t}}$.

For distinct messages $\p{\tilde{w}}$ to be mapped to distinct points $\p{\tilde{t}}$, we require that $\p{GA_S}$ be full rank. 
Since $\p{A_S}$ is full rank, this is same as requiring that $\p{G}$ be full rank. 
Apart from the event $\mathcal{E}\!: Q_{\LS}(\p{z}) \in \LS \backslash \Lc$, we assume that the decoder declares an error whenever the event
\begin{align*}
 \mathcal{G} \! : \rank(\p{G} < K\ell)
\end{align*} 
occurs. Hence, the error probability $\Pp_{\p{S}}$ at the receiver $(\p{S},\sig^2)$ satisfies
\begin{equation} \label{eq:PS_bound_step}
 \Pp_{\p{S}} = \Pp(\mathcal{G} \cup \mathcal{E}) \leq \Pp(\mathcal{G}) + \Pp(\mathcal{E}).
\end{equation}  
From~\eqref{eq:prob_rank_def}, we already know that $\Pp(\mathcal{G})$ is exponentially small in $n$. 
%We now upper bound $\Pp(\mathcal{E})$.

%% We will now use the technique of~\cite[Proposition~1]{OrE_Israel_12} to upper bound $\Pp(\mathcal{E})$. 
Using the given design tolerance $\epsilon$, we set $\delta=2^{\frac{\epsilon}{2}}-1$, which is positive if $\epsilon>0$. Let $\rz=\sqrt{n(1+\delta)\sigz^2}$ be the radius of the typical noise vector and $\ballz=\ball(\p{0},\rz)$. Then,
\begin{align}
 \Pp(\mathcal{E}) &= \Pp(\p{z} \in \ballz)\,\Pp(\mathcal{E} | \p{z} \in \ballz) + \Pp(\p{z} \notin \ballz)\,\Pp(\mathcal{E} | \p{z} \notin \ballz) \nonumber \\
                  &\leq \Pp(\mathcal{E} | \p{z} \in \ballz) + \Pp(\p{z} \notin \ballz). \label{eq:PE_bound_step}
\end{align}
Lemma~\ref{lem:noise_is_small} provides an exponential upper bound on $\Pp(\p{z} \notin \ballz)$. In the following theorem we show that $\Pp(\mathcal{E} | \p{z} \in \ballz)$ is also exponentially small in $n$. The proof of this result uses the technique of~\cite{OrE_IT_16,diP_thesis_14} to bound the number of lattice points lying in an $n$-dimensional ball.

%% For technical reasons, we will assume that the noise standard deviation $\sig$ of every receiver is bounded from below by $\sig_{\min}$, where $\sig_{\min}>0$ is a constant. 
%% R1-EDIT %%
%% Let $\sig_{\min}>0$ be the least noise standard deviation $\sig$ among all the receivers.The knowledge of $\sig_{\min}$ enables us to derive an upper bound on error probability which is independent of the side information matrix $\p{S}$ and the noise standard deviation $\sig$ as long as the condition~\eqref{eq:snr_rate_condition} on the minimum required ${\sf SNR}$ is satisfied. The following result, which follows immediately from~\eqref{eq:sigz}, will be used in the proof of Theorem~\ref{thm:pe_one_receiver},
%% \begin{equation} \label{eq:lower_bound_sigz}
%%  \sigz \geq \sqrt{ \frac{\sig_{\min}^2}{1 + \sig_{\min}^2} }.
%5 \end{equation} 
%% END %

\begin{theorem} \label{thm:pe_one_receiver}
For any receiver $(\p{S},\sig^2)$ with
%\begin{equation*}
 $\frac{1}{2}\log\left( 1+ \frac{1}{\sig^2} \right) > (R+\epsilon)\left( K - \rank(\p{S}) \right)$,
%\end{equation*} 
and for all large enough $n$, 
\begin{equation*} %\label{eq:thm:pe_one_receiver}
\Pp(\mathcal{E} | \p{z} \in \ballz) \leq 2^{-\frac{n\epsilon}{4}} 
\end{equation*} 
when averaged over the ensemble of random lattice codes.
\end{theorem}
\begin{IEEEproof}
From~\eqref{eq:error_event}, we note that the decoder is in error when $\p{z}$ is closer to some coset $\p{t'}+\Lc$, with $\p{t'} \in \LS/\Lc$ and $\p{t'} \neq \p{0}$, than any point in $\Lc$. The number of competing cosets is $|\LS/\Lc \, \backslash \, \{\p{0}\}|=p^{(K-M)\ell}-1$, and we index them using the non-zero vectors $\p{w'} \in \Fp^{(K-M)\ell} \, \backslash \{\p{0}\}$. To each $\p{w'}$, we associate the coset corresponding to the coset leader
\begin{equation} \label{eq:tdash}
 \p{t'} = \left[ \Bc p^{-1} g(\p{GA_Sw'}) \right] \! \mmod \Lc.
\end{equation} 
Since $\p{G}$ is random, the coset leader $\p{t'}$ associated with a given $\p{w'}$ is a random vector.
Given that $\p{z} \in \ballz$ and $\p{0} \in \Lc$, the Euclidean distance between $\p{z}$ and $\Lc$ is at the most $\|\p{z}-\p{0}\|\leq\rz$. Hence, for an error event to occur, there must exist a coset $\p{t'}+\Lc$ at a distance less than $\rz$ from $\p{z}$, i.e., $|(\p{t'} + \Lc) \cap \ball(\p{z},\rz)| \neq 0$. Indexing the cosets by $\p{w'}$, we upper bound $\Pp(\mathcal{E} | \p{z} \in \ballz)$ using~\eqref{eq:main_proof:1} given in the top of the next page.
\begin{figure*}[t!]
\hrule
\begin{align}
 \Pp(\mathcal{E} | \p{z} \in \ballz) &\leq \Pp \left( \, \bigcup_{\p{w'} \in \Fp^{(K-M)\ell} \backslash \{\p{0}\}} \left\{ \, \left\vert (\p{t'} + \Lc) \cap \ball(\p{z},\rz) \right\vert \neq 0 \, \right\} \, \Big\vert \p{z} \in \ballz \right) \nonumber \\[1ex]
&\leq \sum_{\p{w'} \in \Fp^{(K-M)\ell} \backslash \{\p{0}\}} \Pp \Big( \, \left\vert (\p{t'} + \Lc) \cap \ball(\p{z},\rz) \right\vert \neq 0  \, \Big\vert \p{z} \in \ballz \Big) \nonumber \\[1ex]
&\leq \sum_{\p{w'} \in \Fp^{(K-M)\ell} \backslash \{\p{0}\}} \Eb \Big( \left\vert (\p{t'} + \Lc) \cap \ball(\p{z},\rz) \right\vert \, \Big\vert \p{z} \in \ballz \Big). \label{eq:main_proof:1}
\end{align} 
\hrule
\end{figure*}
% where the second inequality follows from union bound, and 
The last inequality in~\eqref{eq:main_proof:1} follows from the observation 
\begin{equation*}
\p{1} \left\{ \, \left\vert (\p{t'} + \Lc) \cap \ball(\p{z},\rz) \right\vert \neq 0 \, \right\} \leq \left\vert (\p{t'} + \Lc) \cap \ball(\p{z},\rz) \right\vert, 
\end{equation*} 
where $\p{1}\{\cdot\}$ is the indicator function.
Note that the expectation operation in~\eqref{eq:main_proof:1} is with respect to the random vector $\p{t'}$ as well as the effective noise $\p{z}$.

The matrix $\p{A_S}$ has full column rank, and hence, $\p{A_Sw'} \neq \p{0}$ for every $\p{w'} \neq \p{0}$. Using~\eqref{eq:tdash} and applying Lemma~\ref{lem:t_is_uniform}, we see that $\p{t'}$ is uniformly distributed in $p^{-1}\Lc/\Lc=p^{-1}\Lc \cap \vor(\Lc)$. 
Further, from Lemma~\ref{lem:crypto} the distribution of $\p{t'}$ is independent of $\p{z}$.
Hence, the probability mass function of $\p{t'}$ equals \mbox{$|\left(p^{-1}\Lc\right)/\Lc|^{-1}=p^{-n}$} over every element of the set $\left(p^{-1}\Lc\right)/\Lc$. Using this result, we further upper bound \mbox{$\Pp(\mathcal{E} | \p{z} \in \ballz)$} as in~\eqref{eq:main_proof:12} in the next page,
\begin{figure*}[t]
\begin{align}
\Pp(\mathcal{E} | \p{z} \in \ballz) & \leq \sum_{\p{w'}} \sum_{\p{a} \in p^{-1}\Lc/\Lc } \Pp(\p{t'}=\p{a}) \, \Eb \left( \left\vert (\p{a} + \Lc) \cap \ball(\p{z},\rz) \right\vert \big\vert \, \p{z} \in \ballz\right) \nonumber \\
&= \sum_{\p{w'}} \sum_{\p{a} \in p^{-1}\Lc/\Lc } p^{-n} \Eb \left( \left\vert (\p{a} + \Lc) \cap \ball(\p{z},\rz) \right\vert \big\vert \p{z} \in \ballz \right) \nonumber \\
&= p^{-n} \sum_{\p{w'}} \sum_{\p{a} \in p^{-1}\Lc/\Lc } \Eb \left( \left\vert (\p{a} + \Lc) \cap \ball(\p{z},\rz) \right\vert \, \big\vert \p{z} \in \ballz \right)
= p^{-n} \sum_{\p{w'}}  \Eb \left( \left\vert p^{-1}\Lc \cap \ball(\p{z},\rz) \right\vert \, \big\vert \p{z} \in \ballz  \right). \label{eq:main_proof:12}
\end{align} 
\hrule
\end{figure*}
where the last equality follows from the fact that the set of cosets $\left\{ \p{a} + \Lc | \p{a} \in p^{-1}\Lc/\Lc \right\}$ form a partition of $p^{-1}\Lc$. 
Since the number of competing $\p{w'}$ in~\eqref{eq:main_proof:12} is less than $p^{(K-M)\ell}$, and $\left\vert p^{-1}\Lc \cap \ball(\p{z},\rz) \right\vert = \left\vert \Lc \cap \ball(p\p{z},p\rz) \right\vert$, we obtain
\begin{align*}
\Pp(\mathcal{E} | \p{z} \in \ballz) \leq p^{-n} p^{(K-M)\ell} \, \Eb \left( \left\vert \Lc \cap \ball(p\p{z},p\rz) \right\vert \big\vert \p{z} \in \ballz \right).
\end{align*}  
Using Lemma~\ref{lem:counting}, we bound the number of lattice points inside the ball $\ball(p\p{z},p\rz)$, and obtain
\begin{align*}
\Pp(\mathcal{E} | \p{z} \in \ballz) \leq p^{-n} p^{(K-M)\ell} \frac{V_n}{\vol(\Lc)} \left( \rcov(\Lc) + p\rz \right)^n. 
\end{align*} 
%% 
%% R1-EDIT %%
{
Using the bounds \mbox{$p^{(K-M)\ell} \leq 2^{nR(K-M)}$}, from~\eqref{eq:ell_over_n_rate}; $\vol(\Lc) \geq V_n \,n^{\frac{n}{2}} \, 2^{-\frac{n\epsilon}{4}}$, from~\eqref{eq:vol_Lc_bound}; $\sigz \leq 2^{-(R+\epsilon)(K-M)}$, from~\eqref{eq:sigz_upper_bound}; and the relations \mbox{$\rcov(\Lc)=\sqrt{n}$}, \mbox{$\rz=\sqrt{n(1+\delta)\sigz^2}$}, and \mbox{$1+\delta=2^{\frac{\epsilon}{2}}$}, we obtain the sequence of equalities and upper bounds leading to~\eqref{eq:main_proof:123} shown in the next page.
\begin{figure*}
\begin{align}
\Pp(\mathcal{E} | \p{z} \in \ballz) 
&\leq p^{-n} \, 2^{nR(K-M)} \frac{V_n}{V_n n^{\frac{n}{2}} \, 2^{-\frac{n\epsilon}{4}}} \left( \sqrt{n} +  p\sqrt{n(1+\delta)\sigz^2}\right)^n \nonumber \\[1ex]
&=  \frac{2^{nR(K-M)} 2^{\frac{n\epsilon}{4}}}{n^{\frac{n}{2}}} \left( \frac{\sqrt{n}}{p} +  \sqrt{n(1+\delta)\sigz^2}\right)^n 
=  \frac{2^{nR(K-M)} 2^{\frac{n\epsilon}{4}}}{n^{\frac{n}{2}}} \left( \frac{\sqrt{n}}{p} +  \sqrt{n(1+\delta)}2^{-(R+\epsilon)(K-M)}\right)^n  \nonumber \\[1ex]
&=  \frac{2^{nR(K-M)} 2^{\frac{n\epsilon}{4}}}{n^{\frac{n}{2}}} n^{\frac{n}{2}} (1+\delta)^{\frac{n}{2}} \left( \frac{1}{p\sqrt{(1+\delta)}} + 2^{-(R+\epsilon)(K-M)} \right)^n  \nonumber \\[1ex]
&\leq {2^{nR(K-M)} 2^{\frac{n\epsilon}{4}} 2^{\frac{n\epsilon}{4}}} \left( \frac{1}{p} + 2^{-(R+\epsilon)(K-M)} \right)^n 
\leq \frac{2^{nR(K-M)} 2^{\frac{n\epsilon}{2}}}{ 2^{n (R+\epsilon) (K-M)}} \left( \frac{1}{p\,2^{(R+\epsilon)(K-M)}} + 1 \right)^n  \nonumber \\[1ex]
&= \frac{2^{nR(K-M)} 2^{\frac{n\epsilon}{2}}}{2^{nR(K-M)} \, 2^{n \epsilon (K-M)}} \left( \frac{1}{p\,2^{(R+\epsilon)(K-M)}} + 1 \right)^n. \label{eq:main_proof:123}
\end{align} 
\hrule
\end{figure*}
Since $K-M \geq 1$, we have $2^{n \epsilon(K-M)} \geq 2^{n \epsilon}$, and hence, the upper bound~\eqref{eq:main_proof:123} can be further relaxed as
\begin{align*}
\Pp(\mathcal{E} | \p{z} \in \ballz) \leq 2^{-\frac{n\epsilon}{2}}\left( \frac{1}{p\,2^{(R+\epsilon)(K-M)}} + 1 \right)^n.
\end{align*}
Using the inequality~\eqref{eq:prime_constraint_2}, which immediately follows from the choice of the prime integer $p$, we have
\begin{align*}
\Pp(\mathcal{E} | \p{z} \in \ballz) \leq 2^{-\frac{n\epsilon}{2}} \, 2^{\frac{n\epsilon}{4}} = 2^{-\frac{n\epsilon}{4}}.
\end{align*} 
Note that this upper bound holds for all sufficiently large values of $n$.
}
\end{IEEEproof}

We will now combine the result of Theorem~\ref{thm:pe_one_receiver} with~\eqref{eq:PS_bound_step} and~\eqref{eq:PE_bound_step}, and upper bound the error probability $\Pp_{\p{S}}$ at the receiver $(\p{S},\sig^2)$ as
\begin{align*}
 \Pp_{\p{S}} \leq \Pp(\mathcal{G}) + \Pp(\p{z} \notin \ballz) + \Pp(\mathcal{E} \, \vert \, \p{z} \in \ballz).
\end{align*}  
Using Theorem~\ref{thm:pe_one_receiver}, Lemma~\ref{lem:noise_is_small} and~\eqref{eq:prob_rank_def}, we obtain
\begin{align}
 \Pp_{\p{S}} \leq 2^{-\frac{n}{2}} + e^{-\frac{n \left( \delta - {\rm ln}(1+\delta) \right)}{2}} + e^{-\frac{n\sig^2\delta^2}{4}} + 2^{-\frac{n\epsilon}{4}}, \label{eq:PS_expanded}
\end{align} 
for sufficiently large $n$.
%% R1-EDIT %%
Let \mbox{$\sig_{\min}>0$} be the least noise standard deviation $\sig$ among the finitely many receivers in the { multicast channel}.
Then we have ${\sig^2\delta^2}/{4} \geq {\sig_{\min}^2\delta^2}/{4}$. Also, $\delta-{\rm ln}(1+\delta)>0$ as long as $\delta=2^{\frac{\epsilon}{2}}-1$ is positive. 
%% END %%
Consequently, the parameter
\begin{align*}
 \varepsilon = \min \left\{ \frac{1}{2}, \log e \, \left( \frac{\delta - {\rm ln}(1+\delta)}{2} \right) , \log e \, \left( \frac{\sig_{\min}^2\delta^2}{4}\right),\frac{\epsilon}{4} \right\}
\end{align*}  
is positive, and the value of each of the terms on the right-hand side of~\eqref{eq:PS_expanded} is at the most $2^{-n\varepsilon}$. Hence the error probability at the receiver $(\p{S},\sig^2)$ can be upper bounded as
\begin{equation} \label{eq:PS_bound_final}
 \Pp_{\p{S}} \leq 4\cdot2^{-n \varepsilon},
\end{equation} 
for all sufficiently large $n$.
%% R1-EDIT %%
We remark that the minimum required value of $n$ for this upper bound to hold depends only $\epsilon$, and is independent of the side information matrix $\p{S}$.
%% END %%

\subsection{Completing the proof of the main theorem} \label{sec:complete_the_proof}

The bound~\eqref{eq:PS_bound_final} shows that the error probability for a fixed side information matrix $\p{S}$, averaged over the random code ensemble, tends to $0$ as the code dimension increases. Hence, there exists a choice of lattice code (which is chosen for the given side information matrix $\p{S}$) with a small error probability at this receiver. We want to prove a slightly stronger result, viz., there exists a lattice code such that the decoding error probability for every possible side information matrix $\p{S}$ is small as long as the receiver ${\sf SNR}$ is large enough.
In order to prove this result, we consider a hypothetical { multicast network} that consists of one receiver for each possible choice of the matrix $\p{S}$.
Note that two distinct values of the matrix $\p{S}$ that have identical row space constitute equivalent receiver side information configurations. 
Hence, it is enough to consider a { multicast channel} that consists of one receiver corresponding to each possible subspace of $\Fp^K$, where the dimension of the subspace can be between $0$ and $K-1$.
A subspace of dimension $M$, $0 \leq M \leq K-1$, can be mapped to an $M \times K$ matrix whose rows form a basis of the subspace. This map embeds the set $\mathcal{S}$ of all non-equivalent choices of side information matrix $\p{S}$ into $\cup_{M=0}^{K-1} \Fp^{M \times K}$, which is the set of all matrices over $\Fp$ with $K$ columns and at the most $K-1$ rows. 
{%%
Hence, the number of receivers $|\mathcal{S}|$ can be upper bounded as
%% R1-EDIT %%
\begin{align} \label{eq:set_S}
|\mathcal{S}| \leq \sum_{M=0}^{K-1} \left\vert \Fp^{M \times K} \right\vert = \sum_{M=0}^{K-1} p^{MK} \leq Kp^{K^2}. %% \leq K \, n^{\beta K^2}.
\end{align} 
}%
%% END %%
We assume that each receiver $(\p{S},\sig^2)$, $\p{S} \in \mathcal{S}$, satisfies the lower bound~\eqref{eq:snr_rate_condition} on ${\sf SNR}$ and outputs an estimated message vector $\p{\hat{w}}(\p{S})$ using its own channel observation. We say that the { multicast} network is in error if any of the receivers commits a decoding error. 
{%%
Using a union bound argument and the upper bounds~\eqref{eq:PS_bound_final} and~\eqref{eq:set_S}, we see that the network error probability $\Pp_{\rm net}$ averaged over the random ensemble of lattice codes satisfies
%% R1-EDIT %%
\begin{align}
\Pp_{\rm net}&=\Pp(\text{network error}) = \Pp\left( \bigcup_{\p{S} \in \mathcal{S}} \, \left\{ \p{\hat{w}}(\p{S}) \neq \p{w} \right\} \right) \nonumber \\
                          &\leq \sum_{\p{S} \in \mathcal{S}} \Pp\left( \left\{ \p{\hat{w}}(\p{S}) \neq \p{w} \right\} \right)
                          = \sum_{\p{S} \in \mathcal{S}} \Pp_{\p{S}} \nonumber \\
                          &\leq 4K \, p^{K^2} \, 2^{-n\varepsilon}, \label{eq:P_network}
\end{align} 
%% END %%
which tends to $0$ as $n$ becomes arbitrarily large. 
}
Hence, for every sufficiently large $n$, there exists a lattice code such that the network error probability is as small as desired. In particular, this implies that there exists a choice of lattice code such that the decoding error probability at every receiver $(\p{S},\sig^2)$, $\p{S} \in \mathcal{S}$, is simultaneously small.
This completes the proof of the main theorem.

\subsection{Corollaries} \label{sec:corollaries}

%% We now state some relevant corollaries of the main theorem, and remark on the applicability of the random lattice coding schemes in certain multi-terminal communication scenarios.

%% We now state and prove some important corollaries and remarks about the main theorem.

%% We now state a few important corollaries of the main theorem.

\subsubsection{Almost all lattice codes are good}

Using standard arguments based on Markov inequality~\cite{ErZ_IEEE_IT_04,For_Allerton_03,OrE_IT_16}, we show that almost all codes from the random lattice code ensemble yield a small error probability. In order to prove this, it is sufficient to show that for almost all lattice codes the network error probability is small over the hypothetical { multicast} channel that consists of one receiver for each possible side information matrix.

For a given dimension $n$, all the lattice codes in the random code ensemble use the same coarse lattice $\Lc$, but differ in the choice of the fine lattice $\La$ and/or the dither vector $\p{d}$. Let 
\begin{equation*}
X(\La,\p{d}) = \Pp(\text{network error} \, \vert \La,\p{d}) 
\end{equation*} 
denote the network error probability for a given choice of $\La,\p{d}$ in the hypothetical { multicast} channel. If $\La$ and $\p{d}$ are chosen randomly, then $X$ is a random variable. From~\eqref{eq:P_network}, we know that the expected value of $X$, which is equal to the average network error rate $\Pp_{\rm net}$, is small. Suppose we want a lower bound on the fraction of random codes with error probability at the most $2^{-\frac{n\varepsilon}{2}}$. 
{%%
Using Markov inequality, we have
\begin{align*}
 \Pp\left( X > 2^{-\frac{n\varepsilon}{2}} \right) \leq \frac{\Eb \left(X\right)}{2^{-\frac{n\varepsilon}{2}}} \leq \frac{4K \, p^{K^2} \, 2^{-n\varepsilon}}{2^{-\frac{n\varepsilon}{2}}} = 4K \, p^{K^2} \, 2^{-\frac{n\varepsilon}{2}}.
\end{align*} 
}%
It follows that, asymptotically in $n$, for almost all choices of the fine lattice $\La$ and dither vector $\p{d}$, the resulting lattice code $(\La-\p{d})/\Lc$ provides an exponentially small error probability in the { multicast} channel, i.e.,
\begin{align*}
 \lim_{n \to \infty} \Pp\left( X \leq 2^{-\frac{n\varepsilon}{2}} \right) = 1.
\end{align*} 

\subsubsection{Goodness in single-user AWGN channel} \label{sec:corollaries:awgn}

Our model of { multicast} channel includes as a special case the single-transmitter single-receiver AWGN channel with no side information at the receiver, i.e., number of messages $K=1$, side information matrix $\p{S}$ is the empty matrix and $\rank(\p{S})=M=0$. The decoder for this receiver uses the $\ell \times \ell$ identity matrix for $\p{A_S}$ and the all zero vector for $\p{v}$, see~\eqref{eq:set_of_solutions_w}. Specializing the main theorem for a single receiver with $M=0$, we immediately deduce that the ensemble of random lattice codes achieves the capacity of the single-user AWGN channel and hence arrive at Corollary~\ref{cor:awgn_channel}.

%% \begin{corollary}
%% \end{corollary}

It is well known that (nested) lattice codes, and lattice constellations in general, can achieve the capacity of the point-to-point AWGN channel~\cite{UrR_IT_98,ErZ_IEEE_IT_04,OrE_IT_16,diP_thesis_14,LiB_IT_14}. Our corollary to the main theorem provides an alternate proof of this result which is based only on simple counting arguments. 

%% Our proof allows the slowest known rate of growth of the prime $p$ as a function of the code dimension $n$. 

The proof technique presented in this paper relies on lattices obtained by applying Construction~A to random linear codes over a large enough prime field $\Fp$. This technique was introduced by Loeliger in~\cite{Loe_IT_97} and used in~\cite{ErZ_IEEE_IT_04,OrE_IT_16,diP_thesis_14} to prove the goodness of lattice codes in AWGN channel. 
Each of these results requires a different choice of the prime $p$ and places different requirements on the characteristics of the coarse lattice $\Lc$. The following are some of the properties that have been used in the literature:
\begin{itemize}
\item \emph{Rogers-good:} the ratio of covering radius $\rcov(\Lc)$ to the effective radius $\reff(\Lc)$ of the lattice must be close to $1$, see~\eqref{eq:rogers}. Such a lattice is also said to be \emph{good for covering}.
\item \emph{MSE-good:} the value of the lattice parameter 
\begin{equation*}
\frac{1}{n \, \vol(\Lc)^{1+\frac{2}{n}}} \int_{\vor(\Lc)} \|\p{x}\|^2 {\rm d}\p{x}, 
\end{equation*} 
known as the \emph{normalized second moment}, is close to \mbox{${1}/{2\pi e}$}, see~\cite{ELZ_IT_05}. Every Rogers-good lattice is also MSE-good, and hence, this is a weaker requirement.
\item \emph{Poltyrev-good:} such a lattice, when used as an infinite constellation, achieves the capacity of an AWGN channel in which the transmitter has no power constraints~\cite{Pol_IT_94,ELZ_IT_05}. These lattices are resilient against additive white Gaussian noise.
\end{itemize} 
%% R1-EDIT %%
The achievability result of~\cite{ErZ_IEEE_IT_04} requires $\Lc$ to be simultaneously Rogers-good and Poltyrev-good, and uses \mbox{$p=2^{nR}$}, i.e., the prime field $\Fb_p$ used for Construction~A varies with the dimension of the lattice code and the size of the field increases exponentially in $n$. %% (exponential in $n$). 
The random code ensemble of~\cite{OrE_IT_16} uses an MMSE-good lattice for $\Lc$, lets $p$ grow as $n^{1.5}$, and can accommodate a wide class of channel noise statistics, including white Gaussian noise. The code construction of~\cite{diP_thesis_14} requires $p$ to be at least $n^{0.5}$, needs no dithering operation, i.e., uses $\p{d}=\p{0}$, but is known to achieve capacity only if ${\sf SNR}>1$.
%
%% In comparison, our proof method allows $p$ to grow as $n^{\beta}$, for any fixed $\beta>0$, and holds for all ${\sf SNR}>0$, while requiring that $\Lc$ be Rogers-good.
{%%
In comparison, our proof method uses a fixed (albeit large) value of $p$ and holds for any ${\sf SNR}>0$, while requiring that $\Lc$ be Rogers-good.
}%

\section{Conclusion} \label{sec:conclusion}

We have showed that lattice codes are optimal for common message broadcast in Gaussian channels where receivers have side information in the form of linear combinations of source messages. We used random lattice ensembles obtained by applying Construction~A to linear codes over appropriately large prime fields $\Fp$.
%% R1-EDIT %%
%% The growth of $p$ as $n^{\beta}$ does not necessarily pose a limitation in communication applications. 
{%% 
The lower bound $\textstyle p \geq \max\left\{ 2^{2KR} , (2^{\frac{\epsilon}{4}}-1)^{-1} 2^{-R}  \right\}$ on the value of $p$ does not necessarily pose a limitation in communication applications. 
For instance, in the relay network of Example~\ref{ex:relay_network}, the first phase of the protocol, namely compute-and-forward~\cite{NaG_IT_11}, only requires that $\frac{n}{p} \to 0$ as $n \to \infty$, which can be met by our scheme by varying $p$ with the dimension $n$: for instance, by choosing $p$ to be the smallest prime greater than or equal to $n^{\beta}$ for a fixed $\beta>1$.
This will also ensure that the inequality $\textstyle p \geq \max\left\{ 2^{2KR} , (2^{\frac{\epsilon}{4}}-1)^{-1} 2^{-R}  \right\}$ holds for all sufficiently large values of $n$.
}% 
Similarly, with Example~\ref{ex:overlay}, where the broadcast signal supplements a wired multicast network, it is known that wireline network codes meeting the $\maxflow$ bound exist over every large enough finite field~\cite{Yeu_Springer_08}. 
{%%
Hence, we can choose $p$ to be sufficiently large to simultaneously optimize both the wired and wireless parts of the hybrid network.
On the other hand, designing lattice strategies for a fixed small size of the finite field, especially sizes that are powers of two, may have greater practical significance.
}%
%% END %%

The capacity of the Gaussian broadcast channel with receiver side information under general message demands, such as with private message requests, is known only for some special cases~\cite{Wu_ISIT_07,AOJ_ISIT_14,SiC_ISIT_14}. 
%% R1-EDIT GaussianR1 %%
The proofs for achievability in these cases utilize ensembles of codebooks generated using the Gaussian distribution together with dirty-paper and superposition coding. 
%% END %%
It will be interesting to examine if the lattice structure of the codes proposed in this paper can be exploited to derive new capacity results beyond the known cases.
 
%% We showed that lattice codes are sufficient to achieve the capacity of Gaussian broadcast channels with coded and uncoded side information at the receivers.
%% \begin{itemize}
%% \item Prime $p$ grows with $n$. This is not a problem for wireline multicasting or with compute-and-forward. How to encode for a fixed finite field alphabet.
%% \item General message demands.
%% \item Code constructions. Can review lattice index coding.
%% \end{itemize} 

%% R1-EDIT %% 
\appendix
{

\section*{Capacity of the Common Message Gaussian Broadcast Channel with Coded Side Information} \label{app:capacity_sketch}

Consider the problem setup with a single transmitter and $N$ receivers $(\p{S}_1,\sig_1^2),\dots,(\p{S}_N,\sig_N^2)$ as described in Section~\ref{sec:channel_model}. We now provide a sketch of the proof that $C$, defined in~\eqref{eq:capacity}, is the capacity of this channel.

\subsection{Converse}

Suppose there exists a coding scheme that achieves rate $R$ in the multicast channel with vanishing decoding error probability at all the receivers. Let the scheme transmit one realization of $(w_1,\dots,w_K)$ for every $\kappa$ channel uses, i.e., $R=\frac{1}{\kappa} \, \log_2 p$. From~\eqref{eq:conditional_entropy} the conditional entropy of each realization of $(w_1,\dots,w_K)$ at the $i^{\text{th}}$ receiver $(\p{S}_i,\sig_i^2)$, given the corresponding coded side information, is $(K-M_i) \log_2 p$. The per-channel use conditional entropy of the message is thus 
\begin{equation*}
\frac{(K-M_i) \log_2 p}{\kappa} = (K-M_i) \, R.
\end{equation*} 
In order to guarantee reliable communication it is necessary that the mutual information between the channel input at the transmitter and the channel output at the $i^{\text{th}}$ receiver be greater than the conditional entropy $(K-M_i)\,R$. Since the input power is constrained to be at the most $1$ and the noise variance at the $i^{\text{th}}$ receiver is $\sig_i^2$, the maximum mutual information is $\frac{1}{2} \log_2 \left( 1 + \frac{1}{\sig_i^2} \right)$, and hence we have
\begin{align*}
 (K-M_i) \, R &< \frac{1}{2} \log_2 \left( 1 + \frac{1}{\sig_i^2} \right), \text{ or equivalently,} \\ R &< \frac{1}{(K-M_i)} \cdot \frac{1}{2} \log_2 \left( 1 + \frac{1}{\sig_i^2} \right).
\end{align*} 
Considering all the $N$ receivers we immediately deduce that $R<C$.

\subsection{Achievability}

The proof of achievability closely follows the proof of Theorem~6 of~\cite{Tun_IEEE_IT_06} and the standard textbook argument used for the achievability of the capacity of single-user AWGN channel.
%% based on an ensemble of codebooks generated using the Gaussian distribution.
%
Let $\epsilon > 0$ be any constant. For a given code length $n$ choose the message length $\ell$ as the largest integer such that the rate $R= \frac{\ell}{n} \, \log_2 p$ satisfies
\begin{equation*}
R  < \min_{i \in \{1,\dots,N\}} \frac{1}{(K-M_i)}\cdot\frac{1}{2}\log_2\left( 1 + \frac{1-\epsilon}{\sig_i^2} \right) - 3\epsilon.
\end{equation*}
As $n \to \infty$, it is straightforward to show that $R$ converges to the right-hand side of the above inequality.
For each of the $2^{nKR}$ message vectors $\p{w}={(\p{w}_1^\intercal,\dots,\p{w}_K^\intercal)}^\intercal \in \Fb_p^{K\ell}$, associate a codeword $\p{x}(\p{w}) \in \Rb^n$ each of whose components are generated independently using the Gaussian distribution with zero mean and variance $(1-\epsilon)$. These $2^{nKR}$ vectors constitute the randomly-generated $n$-dimensional codebook $\mathcal{X}$. 

\subsubsection*{Encoding}
%% The codebook $\mathcal{X}$ is revealed to the transmitter and all the receivers. 
If the source message is $\p{w}$, the transmitter broadcasts the vector $\p{x}(\p{w})$ over $n$ channel uses.

\subsubsection*{Decoding}
Consider the $i^{\text{th}}$ receiver $(\p{S}_i,\sig_i^2)$ that observes the channel output $\p{y}_i$ and the side information 
$\sum_{k=1}^{K}s_{m,k}^{(i)}\p{w}_k=\p{u}_m^{(i)}$, $m=1,\dots,M_i$, where $\p{S}_i=[s_{m,k}^{(i)}]$. 
As in~\eqref{eq:subcode_expurgated}, the receiver determines the subcode $\mathcal{X}_{\rm sub}=\mathcal{X}(\p{S}_i,\p{u}_1^{(i)},\cdots,\p{u}_{M_i}^{(i)})$ of the codebook $\mathcal{X}$ that corresponds to the set of all message vectors $\p{w}$ which are consistent with the observed coded side information.
Among the codewords in $\mathcal{X}_{\rm sub}$, the decoder chooses the vector that is jointly (weakly) $\epsilon$-typical with $\p{y}_i$. If there exists a unique such codeword $\p{x}(\p{\hat{w}})$ that additionally satisfies the power constraint $\|\p{x}(\p{\hat{w}})\|^2 \leq n$, the receiver declares $\p{\hat{w}}$ as the decoded message. Otherwise the receiver declares a decoding error. 

Given that $\mathcal{X}_{\rm sub}$ consists of $2^{n(K-M_i)R}$ vectors generated independently using the Gaussian distribution with zero mean and variance $(1-\epsilon)$ and 
\begin{equation*}
 (K-M_i)\,R < \frac{1}{2}\log_2\left( 1 + \frac{1-\epsilon}{\sig_i^2} \right) - 3\epsilon,
\end{equation*}
it is routine to show that the probability of decoding error at the $i^{\text{th}}$ receiver, averaged over the ensemble of codebooks, decays exponentially with code length $n$~\cite[proof~of~Theorem~10.1.1]{CoT_JohnWiley_91}. It follows that the probability that any of the $N$ receivers commits a decoding error is also exponentially small in $n$. Hence, there exists at least one codebook $\mathcal{X}$ that transmits each message at rate $R$ with the decoding error probability at all the receivers as small as desired. Letting $n \to \infty$ and $\epsilon \to 0$, we observe that any rate $R<C$ is achievable.

}%

\section*{Acknowledgment}
The authors would like to thank the anonymous reviewers whose comments have improved the content and the presentation of this paper.

%%%% references %%%%%

% % Biographies

\vspace*{-20\baselineskip}
\begin{IEEEbiographynophoto}{Lakshmi Natarajan}
is an Assistant Professor in the Department of Electrical Engineering, Indian Institute of Technology Hyderabad. He received the B.E. degree from
the College of Engineering, Guindy, in electronics and communication in 2008, and the Ph.D. degree from the Indian Institute of Science, Bangalore, in 2013. Between 2014 and 2016 he held a post-doctoral position at the Department of Electrical and Computer Systems Engineering, Monash University, Australia. His primary research interests are coding and information theory for communication systems.

Dr. Natarajan is an Editor of the \textsc{IEEE Wireless Communications Letters}. He was the recipient of the Seshagiri-Kaikini Medal 2013-14 for best Ph.D. thesis, Department of Electrical Communication Engineering, Indian Institute of Science, Bangalore. He was recognized as an Exemplary Reviewer by the editorial board of the \textsc{IEEE Wireless Communications Letters} in 2013, 2015 and 2016. He served as the Local Arrangements Co-Chair of the 2016 {\em Australian Communications 
Theory Workshop}, Melbourne and the 2016 {\em Australian Information Theory School}, Melbourne.
\end{IEEEbiographynophoto}

\newpage

\begin{IEEEbiographynophoto}{Yi Hong}(S'00--M'05--SM'10)
is currently a Senior Lecturer at the Department of Electrical and Computer Systems Eng., Monash University, Melbourne, Australia.
She obtained her Ph.D. degree in Electrical Engineering and Telecommunications from the University of New South Wales (UNSW), Sydney, and received the {\em NICTA-ACoRN Earlier Career Researcher Award} at the {\em Australian Communication Theory Workshop}, Adelaide, Australia, 2007. 

Dr. Hong was an Associate Editor for the \textsc{IEEE Wireless Communications Letters} and the {\em Transactions on Emerging Telecommunications Technologies (ETT)}. She was the General Co-Chair of the {\em IEEE Information Theory Workshop} 2014, Hobart; the Technical Program Committee Chair of the {\em Australian Communications Theory Workshop} 2011, Melbourne; and the Publicity Chair at the {\em IEEE Information Theory Workshop} 2009, Sicily. She was a Technical Program Committee member for several leading IEEE conferences. Her research interests include communication theory, coding and information theory with applications to telecommunication engineering.
\end{IEEEbiographynophoto}

\vspace*{-20\baselineskip}

\begin{IEEEbiographynophoto}{Emanuele Viterbo}(M'95--SM'04--F'11)
is currently a Professor in the ECSE Department and an Associate Dean in Graduate Research at Monash University, Melbourne, Australia. He received his Ph.D. in 1995 in Electrical Engineering, from the Politecnico di Torino, Torino, Italy. From 1990 to 1992 he was with the European Patent Office, The Hague, The Netherlands, as a patent examiner in the field of dynamic recording and error-control coding. Between 1995 and 1997 he held a post-doctoral position in the Dipartimento di Elettronica of the Politecnico di Torino. In 1997-98 he was a post-doctoral research fellow in the Information Sciences Research Center of AT\&T Research, Florham Park, NJ, USA. From 1998-2005, he worked as Assistant Professor and then Associate Professor, in Dipartimento di Elettronica at Politecnico di Torino. From 2006-2009, he worked in DEIS at University of Calabria, Italy, as a Full Professor.

Prof.~Viterbo is an ISI Highly Cited Researcher since 2009. He was an Associate Editor of the \textsc{IEEE Transactions on Information Theory}, the \emph{European Transactions on Telecommunications} and the \emph{Journal of Communications and Networks}, and Guest Editor for the \textsc{IEEE Journal of Selected Topics in Signal Processing: Special Issue on Managing Complexity in Multiuser MIMO Systems}. Prof. Viterbo was awarded a NATO Advanced Fellowship in 1997 from the Italian National Research Council. His main research interests are in lattice codes for the Gaussian and fading channels, algebraic coding theory, algebraic space-time coding, digital terrestrial television broadcasting, digital magnetic recording, and irregular sampling.
\end{IEEEbiographynophoto}

\end{document}